# Terahertz Channels in Atmospheric Conditions: Propagation Characteristics and Security Performance

Jianjun Ma[1,2], Yuheng Song[1,2], Mingxia Zhang[1,2], Guohao Liu[1,2], Weiming Li[1,2], John F. Federici[3], Daniel M. Mittleman[4]

[1]School of Integrated Circuits and Electronics, Beijing Institute of Technology, Beijing 100081 China
[2]Beijing Key Laboratory of Millimeter Wave and Terahertz Technology, Beijing 100081 China
[3]Department of Physics, New Jersey Institute of Technology, Newark, NJ 07102 USA
[4]School of Engineering, Brown University, Providence, RI 02912 USA

## Abstract

With the growing demand for higher wireless data rates, the interest in extending the carrier frequency of wireless links to the terahertz (THz) range has significantly increased. For long-distance outdoor wireless communications, THz channels may suffer substantial power loss and security issues due to atmospheric weather effects. It is crucial to assess the impact of weather on high-capacity data transmission to evaluate wireless system link budgets and performance accurately. In this article, we provide an insight into the propagation characteristics of THz channels under atmospheric conditions and the security aspects of THz communication systems in future applications. We conduct a comprehensive survey of our recent research and experimental findings on THz channel transmission and physical layer security, synthesizing and categorizing the state-of-the-art research in this domain. Our analysis encompasses various atmospheric phenomena, including molecular absorption, scattering effects, and turbulence, elucidating their intricate interactions with THz waves and the resultant implications for channel modeling and system design. Furthermore, we investigate the unique security challenges posed by THz communications, examining potential vulnerabilities and proposing novel countermeasures to enhance the resilience of these high-frequency systems against eavesdropping and other security threats. Finally, we discuss the challenges and limitations of such high-frequency wireless communications and provide insights into future research prospects for realizing the 6G vision, emphasizing the need for innovative solutions to overcome the atmospheric hurdles and security concerns in THz communications.

# 1. Introduction

Terahertz (THz) technology, spanning frequencies from 0.1 to 10 THz, has emerged as a frontier in scientific research and technological innovation over the past few decades [1, 2]. The unique properties of THz waves—including their ability to penetrate non-metallic and non-polar materials, sensitivity to molecular vibrations and rotations, and non-ionizing interaction with matter—make them exceptionally attractive for a diverse array of applications [3, 4]. Recent advancements in THz generation and detection have significantly expanded the practical potential of this technology. Notably, the development of quantum cascade lasers operating within the THz spectrum [5], photonics-based THz generation methods [6], and novel detectors based on field-effect transistors and semiconductor technologies [7] have paved the way for integrating THz systems into real-world applications..

THz waves offer unparalleled high-resolution capabilities due to their short wavelengths, which bridge the gap between microwave and infrared frequencies [8, 9]. This attribute is revolutionizing fields such as security screening, non-destructive testing, and autonomous vehicle sensing. For instance, the development of compact THz radar systems with millimeter-scale resolution [10] enables the detection of concealed weapons, identification of chemical compounds, and detailed imaging through clothing and packaging materials [3]. In the automotive industry, THz radar is being explored for high-resolution imaging under adverse weather conditions, effectively complementing existing sensor technologies to enhance vehicular safety and autonomy [11]. Moreover, THz waves have demonstrated strong sensitivity to water content and biomolecular structures, making them highly suitable for non-invasive diagnostics and imaging in the biomedical domain [12, 13]. THz spectroscopy and imaging techniques have been applied to analyze DNA, proteins, and living cells, providing deeper insights into biological processes at the molecular level [14-17]. The development of calibration-free THz sensors for detecting cancerous changes in gastric cells, underscores the significant role of THz technology in enhancing early-stage cancer detection and improving diagnostic accuracy [18]. Furthermore, groundbreaking studies have shown that the myelin sheath functions as a dielectric waveguide for signal propagation in the THz range [19], and that THz waves can enhance the permeability of voltage-gated calcium channels [20], opening new avenues for understanding neural information transmission and influencing cellular mechanisms in biomedical applications.

As the demand for higher data rates and lower latencies escalates with the advent of emerging applications, wireless communication in the THz range has surfaced as a promising key technology for the forthcoming era of 6G [21]. THz communications represent a paradigm shift, offering unprecedented capabilities that address the limitations of current radio frequency (RF) and millimeter-wave (mm-Wave) systems. While 5G systems primarily operate within the mm-Wave band and are approaching theoretical limits in spectral efficiency and capacity [22], the THz spectrum provides vast bandwidth resources capable of supporting multi-gigabit per second (Gbps) or even terabit per second (Tbps) data rates [23]. These capabilities are essential for achieving the ambitious performance metrics envisioned for 6G networks, including peak data rates of 1 Tbps, user-experienced data rates of 10 Gbps, and ultra-low latencies of 0.1 ms [24, 25].

**1.1 Terahertz communications**

The technological breakthroughs, as we mentioned above, have paved the way for a wide array of potential applications of THz communication (Fig. 1), such as 1) ultra-high-speed cellular systems [21] for Tbps-level connections, revolutionizing mobile broadband experiences; 2) Terabit wireless local area networks (Tera-WiFi) [26] for enhancing indoor wireless connectivity dramatically; 3) Terabit IoT (Tera-IoT) [27], facilitating seamless connectivity among countless devices, addressing the growing demands of IoT applications; 4) Terabit Integrated Access and Backhaul (Tera-IAB) [28], enhancing network capacity and flexibility in wireless backhaul solutions; 5) Ultra-wideband THz space communications (Tera-SpaceCom) [29], offering exciting prospects for satellite and deep space communications; and 6) Wireless Network-on-Chip (WiNoC) [30], enabling high-speed, low-latency interconnects, potentially revolutionizing computer architecture. Beyond these, the THz band also offers unique capabilities for integrated sensing and communications (ISAC), which enables novel applications such as high-precision virtual/augmented reality experiences [31], advanced vehicular communications and radar sensing [11], and millimeter-level indoor positioning [32]. Additionally, the impact of THz waves on biological ion channels has shown significant implications for enhancing signal permeability and selectivity, which could be leveraged for innovative THz intra-body communication and sensing applications [20].

The small wavelength of THz signals (0.03 to 3 mm) enables the development of compact, highly integrated communication systems [33]. This characteristic is particularly advantageous for applications where space is at a premium, such as in satellite communications or personal mobile devices. Furthermore, THz waves can exhibit strong directionality, which can be exploited to enhance communication security at physical layer and enable spatial multiplexing techniques [1, 34].

The potential of THz communications has garnered worldwide attention. In 2019, the U.S. Federal Communications Commission (FCC) created a new category of experimental licenses and allocated over 20 GHz of unlicensed spectrum between 95 GHz and 3 THz to facilitate testing for 6G and beyond technologies [35]. Similarly, the European Horizon

2020 program and key projects funded by the Chinese Ministry of Science and Technology have been actively supporting THz communication research and development [36, 37]. However, despite these advantages and ongoing research efforts, THz communications face significant challenges that must be addressed before widespread deployment can be achieved. One of the most critical of these is the impact of atmospheric conditions on THz channel propagation. THz waves are highly susceptible to molecular absorption, particularly by water vapor, as well as scattering by atmospheric particles such as rain or snow, and also by atmospheric turbulence [38]. These effects can cause severe attenuation and signal distortion, potentially limiting the effective range and reliability of THz channels. Furthermore, the unique propagation characteristics of THz waves necessitate the development of new channel models, adaptive modulation and coding schemes, and network planning methodologies [39-41]. The design of THz transceivers and antennas must also evolve to mitigate channel impairments and exploit the unique properties of THz propagation [42].

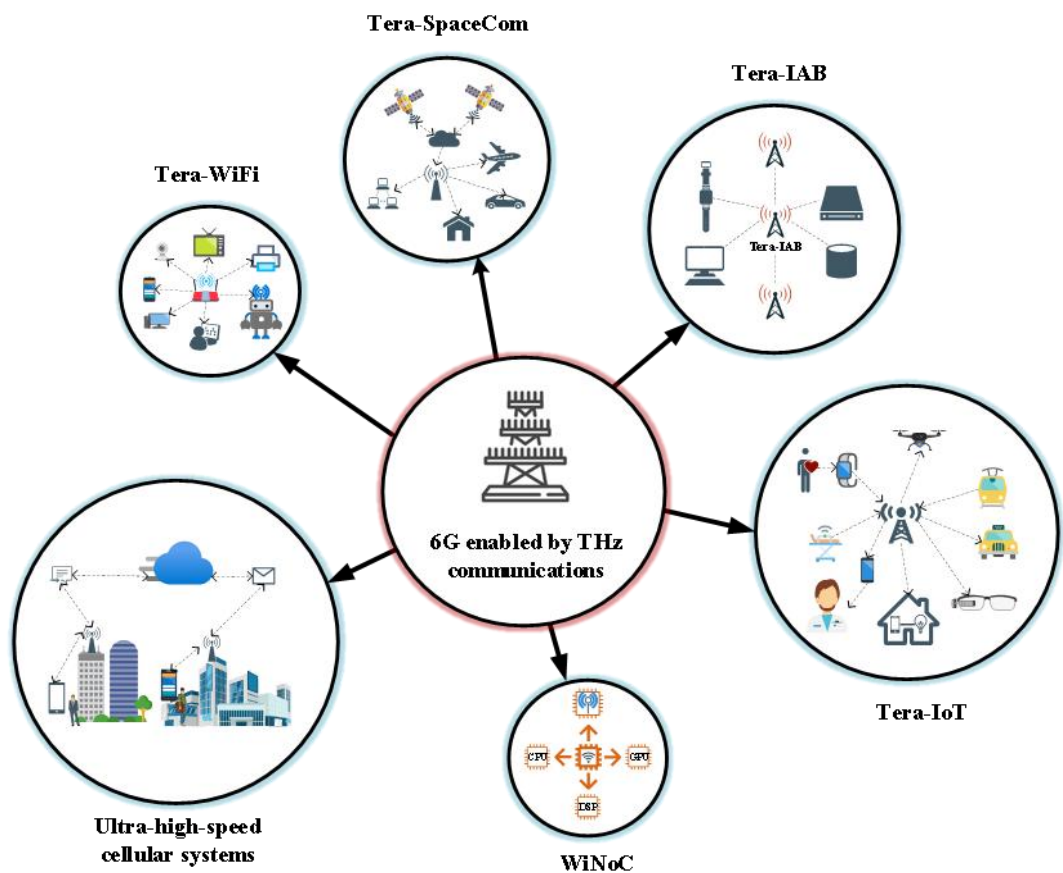

Figure 1 6G enabled by THz communications

### 1.2 Weather impact

Unlike lower frequency bands, THz waves are highly susceptible to various atmospheric phenomena, which can dramatically affect the performance and reliability of THz channels. Understanding these effects is crucial for accurate channel modeling, link budget optimization, and the development of robust communication protocols. The fundamental propagation characteristics of THz channels can be explored using the Friis transmission formula [43-45], as

$$P_{out} = P_{in} \left( \frac{\lambda}{4\pi d} \right)^2 G_r \cdot G_t \cdot F_r(\theta_r, \phi_r) \cdot F_t(\theta_t, \phi_t) \cdot e^{-\alpha_e d} \cdot \varepsilon_p \qquad (1)$$

with $P_{in}$ being the input power to the transmitting antenna, $\lambda$ as the wavelength of the radiation, and $d$ as the distance between the transmitting ($t$) and receiving ($r$) antenna [46], The parameter $G_{r/t}$ is the antenna gain on the receiver/transmitter side, $F_{r/t}$ is the normalized intensity pattern function, $\varepsilon_p$ is the polarization coupling efficiency. The term $e^{-\alpha_e d}$ represents the channel power transmission factor with $\alpha_e$ being the extinction (or total attenuation) coefficient due to both scattering and absorption effects in the atmosphere. This formula highlights several key factors affected by atmospheric conditions, particularly the extinction coefficient $\alpha_e$. The impact of weather conditions on THz channels is primarily due to variations in the atmospheric complex refractive index, which directly influence key propagation parameters such as absorption, scattering and scintillation [47]. These changes consequently affect the overall performance of the THz communication channels (see Fig. 2), as

- Link budget: Weather-induced attenuation directly affects the link budget, potentially reducing the effective communication range and reliability.

- Frequency selectivity: The strong frequency dependence of atmospheric effects, particularly absorption, creates a highly frequency-selective channel. This necessitates adaptive modulation and coding schemes to maintain optimal performance.

- Temporal variations: Weather conditions can change rapidly, leading to time-varying channel characteristics. This requires dynamic adaptation of communication parameters.

- Spatial effects: Phenomena like beam wandering and scattering can affect the spatial characteristics of the THz channel, impacting beamforming and spatial multiplexing techniques.

- Security implications: Weather-induced scattering can create multipath propagation scenarios, potentially increasing vulnerability to eavesdropping attacks. Conversely, the strong absorption by water vapor can be exploited to create secure, atmosphere-limited line-of-sight links.

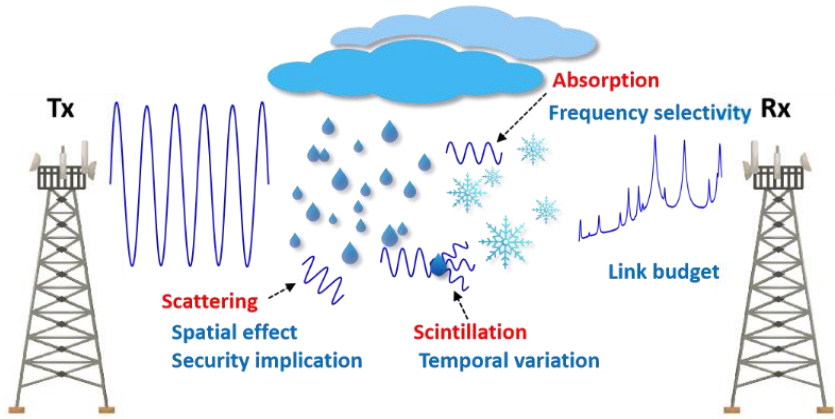

**Figure 2 Impact of atmospheric effect on THz communications**

Understanding and mitigating these weather impacts is crucial for the practical deployment of THz communication systems. This requires accurate channel modeling that incorporates these atmospheric effects, as well as the development of adaptive techniques to maintain link performance under varying weather conditions. Furthermore, the unique security implications of weather effects on THz channel propagation must be thoroughly understood and addressed to ensure the confidentiality and integrity of THz communications [43].

### 1.3 Purpose of this article

The primary purpose of this review article is to provide a comprehensive analysis of the propagation characteristics and security properties of THz channels under various atmospheric conditions. By synthesizing our recent research and experimental findings, we aim to offer insights into the challenges and potential solutions for THz communication systems. The review will cover fundamental principles, measurement techniques, and modeling approaches for THz channel performance, followed by an in-depth examination of the impact of different weather conditions on THz propagation and security.

## 2. Fundamental of Terahertz Channel Performance in Atmospheric Conditions

THz frequency range is particularly susceptible to various atmospheric phenomena, including molecular absorption, scattering by particles, and turbulence-induced scintillation effects. These factors can significantly impact the performance, reliability, and propagation range of THz links. Therefore, a comprehensive understanding of THz channel behavior under different atmospheric conditions is essential for accurate system design, performance prediction, and the development of effective mitigation strategies. Understanding the fundamental characteristics, measurement techniques, and modeling approaches for THz channels in atmospheric conditions is crucial for better navigating the complexities of THz propagation in real-world environments and develop more effective solutions for future THz communication systems.

## 2.1 Channel characteristics

The propagation of THz channels through the atmosphere is significantly influenced by various physical phenomena, including absorption, scattering, and refraction. These effects collectively shape the fundamental channel characteristics and ultimately determine the performance of THz communication systems in real-world scenarios.

**Absorption**

THz waves are strongly absorbed by atmospheric gases, particularly water vapor. This absorption is highly frequency-dependent, creating distinct transmission windows and attenuation peaks across the THz spectrum. Yang et al. [47, 48] performed detailed measurements of water vapor absorption in the THz range using THz time-domain spectroscopy (THz-TDS) and found that the absorption coefficient of water vapor can exceed 100 dB/km at certain frequencies, creating severe limitations for long-distance THz communications. However, they also identified several transmission windows with relatively low attenuation, such as those centered around 380 GHz, 448 GHz, and 620 GHz. The absorption coefficient can be calculated by summing the effects of individual molecular absorption lines, which can be obtained from spectroscopic databases like HITRAN (High-resolution Transmission molecular absorption) [49, 50]. Siles et al. [51] developed a comprehensive model for atmospheric attenuation in the THz band, incorporating the effects of both water vapor and oxygen. Their model, validated against experimental data, provides a valuable tool for predicting the performance of THz channels under various atmospheric conditions. A line-by-line calculation, including also the continuum contribution, is provided by the ITU Recommendation Sector (ITU-R) [52] base on the physical model MPM93 [53]. This method is proved to be valid at frequencies between 1 and 450 GHz, although it is in less good agreement with measurements at higher frequencies [49]. The results shown in Fig. 3 suggest the importance of the atmospheric absorption contribution to the total attenuation, even at a frequency of 300 GHz which is far from a water vapor absorption resonance. We see that, even when the operating frequency lies in one of the transmission windows, the humidity can affect the channel attenuation significantly. As shown in Fig. 3, when relative humidity changes from 30% to 90%, the path loss (in dB/km) at 300 GHz increases from 143 to 146.

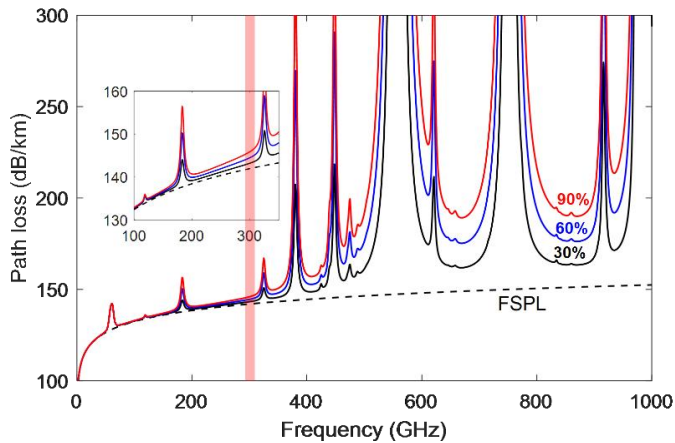

**Figure 3 Channel path loss in free space. (T = 0ºC, P=1013 hPa and RH = 30%, 60% and 90%)**

In addition to attenuation, molecular absorption in the atmosphere also leads to significant group velocity dispersion (GVD) for broadband THz channels. This dispersive effect arises from the frequency-dependent refractive index associated with the molecular absorption lines. Strecker et al. [54, 55] demonstrated that GVD can severely impact the performance of high-bandwidth THz channels, even when operating in atmospheric transmission windows. Their analysis showed that for a 30 GBd (60 Gbps) channel at 250 GHz, GVD limits the maximum transmission distance to approximately 9 km before intersymbol interference becomes prohibitive. At higher data rates, this dispersion limit becomes even more restrictive. Based on their theoretical framework for predicting symbol error rates, they also revealed that in some scenarios, GVD rather than attenuation becomes the primary factor limiting channel performance.

**Scattering**

Scattering occurs when THz waves interact with particles in the atmosphere. The scattering mechanism depends on the ratio of particle size to THz wavelength. Rayleigh scattering occurs when particles are much smaller than the wavelength and Mie scattering occurs when particle size is comparable to the wavelength. For THz waves, Mie scattering is often the dominant mechanism for interaction with rain, snow droplets and larger particles. We conducted a comprehensive study on the scattering effects of rain on THz waves [56-58] and found that for frequencies above 100 GHz, the specific attenuation due to rain increases rapidly with rain rate. The results showed that at 625 GHz, a rain rate of 100 mm/h can lead to an attenuation of about 170 dB/km, which agrees with our theoretical predictions [57] with the scattering cross-section calculated using Mie theory, considering factors such as particle size distribution and complex refractive index.

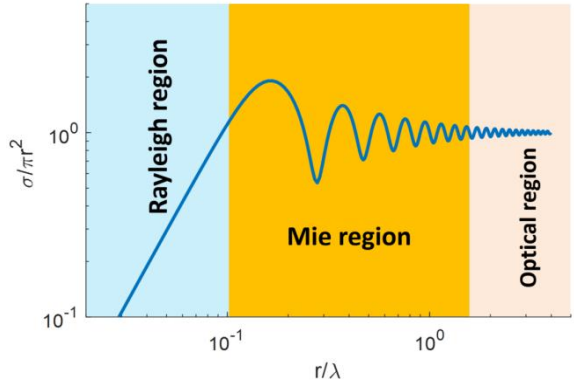

**Figure 4 Rayleigh and Mie scattering.**

**Refractive effects (scintillation)**

Atmospheric turbulence, characterized by random fluctuations in the refractive index of air, leads to scintillation effects in THz channel propagation. These effects manifest as intensity fluctuations, phase front distortions, and beam wandering [59]. Scintillation can significantly impact the reliability and quality of THz channels, particularly over long distances [60]. The severity of turbulence is typically quantified using the refractive index structure parameter $C_n^2$, which ranges from $10^{-17}$ m$^{-2/3}$ for weak turbulence to $10^{-13}$ m$^{-2/3}$ for strong turbulence [61]. To measure the intensity fluctuations caused by scintillation, researchers often use the scintillation index $\sigma_I^2$, which is related to the $C_n^2$ through the equation as $\sigma_I^2 \approx 23.17\ C_n^2\ k^{7/6}\ L^{11/6}$, where $k$ represents the wave number and $L$ is the propagation distance. We conducted experiments comparing THz and infrared (1550 nm wavelength) channel performance in controlled atmospheric turbulence, demonstrating that THz channels are generally less affected by turbulence-induced scintillation than infrared. This is due to the longer wavelength of the THz channels, which makes it less susceptible to small-scale refractive index fluctuations, compared to infrared [62].

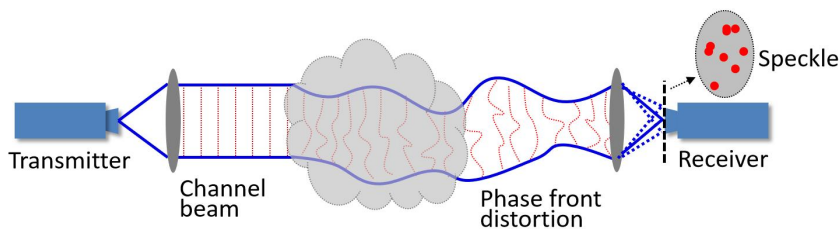

**Figure 5 Atmospheric turbulence - induced phase front distortion and power fluctuation at receiver**

When modeling THz channel characteristics for various weather conditions, it's essential to consider the specific phenomena dominant in each scenario. In clear weather, the primary factors are molecular absorption by water vapor, which create distinct transmission windows and attenuation peaks across the THz spectrum [38, 43]. Additionally, clear atmospheric turbulence leads to scintillation effects, causing intensity fluctuations and phase distortions [62, 63]. For rainy conditions, both absorption and Mie scattering become significant, with specific attenuation increasing rapidly with rain rate and frequency [56, 58, 64]. Snow presents unique challenges due to the complex shapes and varying densities of snowflakes, often resulting in more severe attenuation than rain at the same precipitation rate [50, 65].

Obviously, atmospheric turbulence effects, characterized by the refractive index structure parameter $C_n^2$, are present in all weather conditions but can be exacerbated by temperature gradients and wind [52]. While THz channels are generally less affected by turbulence than infrared, these effects can still significantly impact channels performance, especially over long distances or in strong turbulence conditions [66]. Models for each weather scenario should incorporate the relevant parameters, such as turbulence strength indicators [40].

## 2.2 Channel measurement techniques

Channel measurements are crucial for understanding and modeling the propagation characteristics of THz channels in diverse environments. They provide essential information on power loss, time delay, and other critical parameters necessary for developing robust THz communication systems. There have been several techniques employed to measure wireless channels in atmospheric conditions.

**THz time-domain spectroscopy**

A typical THz time-domain spectroscopy (THz-TDS) system consists of a femtosecond laser, beam splitter, THz emitter, delay line, and THz detector [67]. THz-TDS uses ultrashort laser pulses to generate and detect broadband THz pulses. It provides both amplitude and phase information, allowing for direct measurement of the complex refractive index. THz-TDS is particularly useful for studying atmospheric absorption and material properties. The key advantage is its coherent detection scheme, which allows for the measurement of the THz electric field as a function of time [64, 68].

THz-TDS typically covers a frequency range from 0.1 to 7 THz in a single measurement, allowing for comprehensive spectral analysis of atmospheric effects [69]. THz-TDS systems can achieve dynamic ranges exceeding 90 dB, enabling the detection of weak signals and the measurement of strong absorption features [70]. However, THz-TDS faces some limitations in atmospheric channel measurements. The mechanical delay line used in most THz-TDS systems limits the measurement speed, making it challenging to study rapidly changing atmospheric conditions [71]. They are also sensitive to environmental fluctuations, which can affect measurement stability in outdoor settings [72].

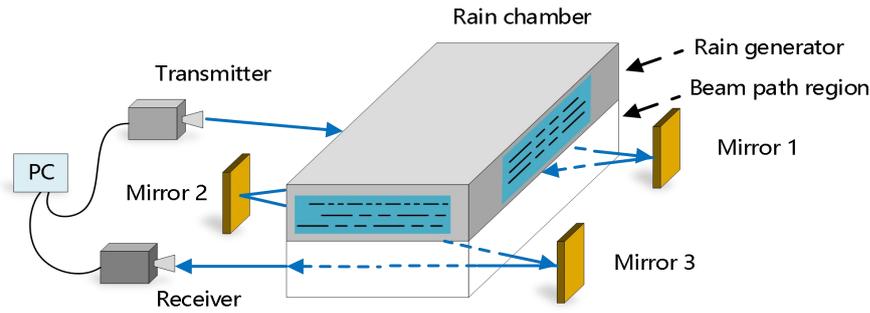

**Figure 6 Schematic measurement setup.**

**Source: Reprinted from [56] with the permission of Elsevier.**

**Vector network analyzer**

Vector network analyzer (VNA) measures the magnitude and phase of scattered (S) parameters, which describe how RF energy propagates through a multi-port network [73]. In THz channel measurements, $S_{21}$ is of particular interest as it represents the channel transfer function. To extend VNA measurements to the THz range, frequency extender modules are used. These modules typically employ harmonic mixers to up-convert the VNA's native frequency range to the desired THz frequency [74]. Priebe et al. [75] used a VNA-based system to characterize indoor THz channels at 300 GHz. Their measurements provided detailed insights into multipath propagation and frequency-selective fading in THz indoor environments. A recent study by Taleb et al. [76] demonstrated the use of a VNA system with WR-2.2 VNAX frequency extension modules to measure THz channel propagation under various atmospheric conditions. The setup achieved a frequency resolution of 100 MHz over a 2 GHz range centered around absorption lines at 380.2 GHz and 448.0 GHz, allowing for detailed characterization of atmospheric effects on THz channels.

VNAs can offer a dynamic range of over 100 dB, allowing for the measurement of highly attenuating channels and/or over long distances. However, VNA-based THz channel measurements face challenges, such as limited output power, increased phase noise at higher frequencies, and potentially long measurement times for wide frequency ranges [77, 78]. Besides, the use of frequency extender modules and harmonic mixers adds complexity to the measurement setup, potentially introducing additional sources of error and requiring careful calibration and alignment.

**Frequency-domain spectral methods**

These methods use continuous-wave (CW) THz sources and detectors to measure channel properties at specific frequencies. They can offer high spectral resolution and are useful for studying narrow atmospheric features. In frequency-domain spectroscopy, the THz signal is generated at specific frequencies and detected using frequency-selective detectors [79]. The primary components of a typical frequency-domain THz system include a CW

THz Source, which is often based on photomixing, quantum cascade lasers (QCLs), or frequency multiplication [80] and a frequency-selective detector, such as Schottky diode mixers or bolometers [81]. Frequency-domain methods have been employed in various atmospheric studies. Norouzian et al. [82] used a frequency-domain system to investigate the attenuation effects of clouds and fog on THz wave propagation, demonstrating the potential of certain THz frequencies for communication through these conditions. We did most of our work on channel measurement in different weather conditions by using such kind of setups.

Frequency-domain systems can achieve high spectral resolution, often sub-MHz, which allows for much detailed characterization of atmospheric absorption features. For narrow-band measurements, these methods can achieve higher signal-to-noise ratio (SNR) compared to time-domain techniques, as the energy is concentrated at specific frequencies. The CW operation allows for simpler signal processing and can be advantageous for communication system testing. However, frequency-domain methods do not provide direct time-domain information, unlike THz-TDS, which can be important for studying multi-path effects. Additionally, sweeping through frequencies can be time-consuming, especially when covering a wide bandwidth. Furthermore, the sensitivity of frequency-selective detectors, such as Schottky diode mixers or bolometers, can vary, affecting the accuracy and reliability of the measurements under different conditions.

**Channel sounders**

The sliding correlator technique is based on the transmission of a pseudo-noise (PN) sequence, typically a maximum length sequence (m-sequence), which has properties similar to white noise but is deterministic [83]. This method allows for high time resolution measurements with reasonable hardware requirements. Lyu et al. [84] developed a THz channel sounder based on the sliding correlator principle for an 84.5 m THz channel at W band. They achieved a time resolution of 0.5 ns and a dynamic range of 40 dB. Prokscha et al. [85] used a sliding correlator channel sounder to investigate the impact of human blockage on THz channels at 300 GHz, providing important data for modeling human body shadowing effects in future THz communication systems.

Although sliding correlator channel sounders can achieve reasonable dynamic ranges, they are often not as high as those achievable with other methods such as VNAs, limiting their effectiveness in highly attenuating environments or over long distances. Extending the sliding correlator technique to THz frequencies requires high-speed electronics and efficient frequency converters, and phase noise in local oscillators at THz frequencies can significantly impact measurement accuracy. Precise synchronization between the transmitter and receiver is crucial, particularly for outdoor measurements. While the sliding correlator technique is faster than swept-frequency techniques, it still requires a finite amount of time to measure the channel, which may pose issues in rapidly changing environments.

**Table 1 Comparison of THz channel measurement methods in different atmospheric conditions**

| Measurement technique | Clear weather | Rain | Snow | Atmospheric turbulence |
|---|---|---|---|---|
| TDS | High dynamic range allows detailed atmospheric characterization. | Capture rapid changes over a wide spectral range. | Can measure complex refractive index changes; Sensitive to environmental fluctuations. | Effective for studying scintillation effects and phase changes; Sensitive to rapid atmospheric changes. |
| VNA | Detailed multipath analysis. | Good for measuring overall attenuation; Struggle with rapid fluctuations. | Good for long-term average measurements; Miss rapid fluctuations due to snowflakes. | Effective for narrow-band turbulence studies; Phase noise and synchronization challenges. |
| FDS | High spectral resolution for narrow-band studies. | Can measure specific frequency attenuation well; Miss rapid temporal changes. | Effective for studying specific frequency attenuation; Less suitable for temporal variations. | Effective for narrow-band turbulence studies; Lack of time-domain information. |
| Channel Sounders | High time resolution for multipath characterization. | Can capture time-varying channel characteristics; Good for studying rain-induced fading. | Capture time-varying effects of snowflakes; Good for studying intermittent blockages. | High time resolution for rapid channel variations; Suitable for detailed multipath analysis. |

For each weather condition - clear weather, rain, snow, and atmospheric turbulence - specific channel measurement methods should be employed to accurately characterize THz channel propagation, as shown in Table 1. In clear weather, THz-TDS is highly effective due to its ability to provide both amplitude and phase information, allowing for detailed atmospheric absorption studies and material property measurements [48, 64]. For rain and snow conditions, a combination of techniques is often necessary to capture both the overall attenuation and rapid fluctuations. Frequency-domain spectral methods are useful for studying specific frequency attenuation in rain and snow [57, 58], while channel sounders can capture time-varying effects and fading characteristics. THz-TDS remains valuable in these conditions for its ability to measure complex refractive index changes, though it can be sensitive to environmental fluctuations [56]. In atmospheric turbulence, THz-TDS and channel sounders are particularly effective due to their ability to study scintillation effects and rapid channel variations. VNA and frequency-domain methods can complement these techniques for narrow-band turbulence studies, though they may face challenges with phase noise and synchronization in rapidly changing environments. The choice of measurement technique should be tailored to the specific aspects of the weather condition being studied, often requiring a multi-method approach to fully characterize the complex THz channel behavior in diverse atmospheric conditions.

**2.3 Channel modeling approaches**

Modeling THz channels in atmospheric conditions is crucial for understanding propagation characteristics and designing effective communication systems. There have been three main approaches commonly used for THz channels modeling [24].

**Deterministic models**

Deterministic models are based on solving Maxwell's equations or using ray-tracing techniques. They can provide accurate results but are computationally intensive and require detailed knowledge of the environment. Examples include finite-difference time-domain (FDTD) simulations for small-scale scenarios and ray-tracing for larger environments. One important deterministic approach is the calculation of absorption lines for atmospheric constituents, particularly water vapor, by using the HITRAN database in conjunction with the Van Vleck-Weisskopf line function [43, 51, 86, 87]. This method allows for accurate modeling of molecular absorption, which is crucial in THz atmospheric propagation, as we demonstrated above. The Mie and Rayleigh scattering models are also deterministic models based on electromagnetic theory [48]. They describe the interaction of electromagnetic waves with particles, taking into account the size, shape, and dielectric properties of the particles relative to the wavelength of the incident radiation. While deterministic methods can offer high accuracy, they are computationally intensive and require detailed knowledge of the environment, which is often difficult to obtain for atmospheric conditions [88].

**Statistical models**

Statistical models characterize the channel using probability distributions derived from measurements or theoretical considerations. They are more tractable for system-level simulations, such as Log-normal distribution for modeling weak turbulence effects, Gamma-gamma distribution for moderate to strong turbulence, Rice or Nakagami distributions for modeling multipath fading [89]. For THz channels in atmospheric conditions, these models often incorporate Rician or Rayleigh fading statistics to model multipath effects [90], Log-normal distributions to model shadowing effects from atmospheric particles [91] and Gamma distributions to model rain attenuation [92]. Markov chain models are also used to model the time-varying nature of THz channels in dynamic atmospheric conditions, which can capture the transitions between different channel states (e.g., clear, light rain, heavy rain) [93]. Other statistical models of atmospheric turbulence, such as the Rytov approximation or the Andrews model, are used to characterize scintillation effects in THz channels [94].Some empirical and semi-empirical models are based on fitting statistical distributions to measured data. Examples include the Crane model for rain attenuation [95], and the ITU-R models, which provide statistical predictions of attenuation based on meteorological parameters [96, 97].

**Hybrid models**

Hybrid methods combine elements of both deterministic and statistical approaches, aiming to balance accuracy and computational efficiency. These methods are gaining popularity in THz channel modeling, especially for atmospheric conditions where both deterministic effects (like absorption by atmospheric gases) and statistical effects (like scattering by particles) play important roles [64]. For instance, a hybrid approach might use deterministic methods to model gaseous absorption while employing statistical methods to account for scattering by atmospheric particles. Another example is the use of the Mie and Rayleigh scattering models in conjunction with statistical methods in THz channel modeling. They are often applied to atmospheric particles whose size distribution is described statistically. For example, the size distribution of raindrops is often modeled using statistical distributions like the Marshall-Palmer distribution [50]. Besides, when modeling a THz channel with many scattering particles, researchers can often use Monte Carlo methods to simulate the collective effect of these particles. This involves repeatedly applying the Mie or Rayleigh scattering calculations to particles sampled from a statistical size distribution [98].

**Table 2 Comparison of THz channel modeling methods in different weather conditions**

| Modeling method | Clear weather | Rain & Snow | Atmospheric turbulence |
|---|---|---|---|
| **Deterministic** | High accuracy on absorption lines; Detailed environmental knowledge; Computationally intensive. | High precision; Computationally intensive; Requires detailed raindrop data; High accuracy for particle interaction. | High precision; Modelling refractive effects accurately; Computationally intensive; Requires detailed turbulence data. |
| **Statistical** | Captures variability; Less computationally intensive; Less precise. | Captures variability in fallrate and attenuation; Effective for system-level simulations; Less precise; Limited by distribution assumptions. | Captures variability in refractive index fluctuations; Effective for dynamic conditions Less precise; Dependent on probability distributions. |
| **Hybrid** | Balances accuracy and efficiency; Complex integration of models; Computationally demanding. | | |

Modeling THz channels in atmospheric conditions presents unique challenges due to the complex interactions between THz waves and atmospheric constituents. The dynamic nature of the atmosphere, with its constantly changing temperature, humidity, and particulate content, makes it particularly challenging to apply deterministic methods effectively [38, 43] In practice, the choice of modeling method depends on the specific application, available computational resources, and the level of accuracy required (Table 2).

For clear weather conditions, deterministic models can provide accurate results, as the propagation environment is relatively stable [24]. However, statistical models like the log-normal distribution are often employed to account for weak turbulence effects that may still be present [90]. In rainy conditions, a hybrid approach is typically most effective. Deterministic methods, like the Mie scattering model, can be used to calculate attenuation due to individual raindrops, while statistical models, such as the gamma distribution, are employed to represent the overall rain attenuation [92]. The Crane model [95] and ITU-R models [96, 97] are examples of empirical models specifically developed for rain and snow attenuation prediction. For snowy conditions, a similar hybrid approach is often used. The Mie scattering model can be applied to individual snowflakes, while statistical distributions are used to model the size and shape variations of snow particles [50]. In the case of atmospheric turbulence, statistical models are predominantly used due to the random nature of turbulence. The Rytov approximation and Andrews model are commonly employed to characterize scintillation effect induced attenuation [94]. Recent advancements in machine learning techniques are being explored to capture complex, non-linear relationships in channel data that may be difficult to model with traditional approaches [99, 100]. Machine learning algorithms, such as k-nearest neighbors (KNN) and Random Forest, have demonstrated high prediction accuracy and computational efficiency in estimating path loss models for air-to-air scenarios [101]. Additionally, artificial neural networks (ANNs) have shown potential in reducing complexity for carrier frequency offset estimation and timing estimation [102]. These efforts show promise in improving the accuracy of THz channel modeling across various atmospheric conditions.

## 3. Impact of Weather Conditions on Channel Propagation

As discussed earlier, THz channel propagation is significantly influenced by atmospheric conditions, primarily due to molecular absorption. While water vapor is the dominant one in the THz band, other atmospheric constituents like oxygen, carbon dioxide, and nitric oxide also play roles, albeit minor in comparison [103]. Building on this understanding of atmospheric effects, we will now examine in detail how specific weather conditions - namely rain, snow, and atmospheric turbulence - impact THz channel performance.

### 3.1 Rain

Rain forms through the precipitation of water vapor in the atmosphere, occurring through different processes in warm and cold clouds. In warm clouds, raindrops form via condensation, collision, and coalescence, starting with tiny droplets of approximately 10 μm radius. Larger raindrops are created as smaller droplets collide and merge. In cold clouds, located above the zero-degree isotherm, rain formation involves ice crystals and super-cooled water droplets, with ice crystals growing by riming and aggregation, producing a wide range of particle sizes that melt into larger raindrops upon descent. Raindrops exhibit varying shapes and sizes. Due to surface tension, a stationary raindrop tends to be spherical. However, as a raindrop falls, pressure decreases at the top and sides and increases at the bottom, causing it to deform into an oblate shape. This deformation becomes significant for raindrops larger than 2.0 mm in diameter. The raindrop size distribution

(DSD) is defined as the number concentration of raindrops with a given diameter $D$ in a specified volume, denoted as $N(D)$ with units $m^{-3}mm^{-1}$. DSD is crucial for calculating rain attenuation and rain rate. Various models describe DSD, including exponential, log-normal, gamma, and normalized gamma distributions. Laws and Parsons initially proposed an empirical equation for DSD [104], later refined by Marshall and Palmer through measurements of raindrops on dyed filter paper [105], as

$$N(D) = N_0 \exp(-\Lambda D) \qquad (1)$$

where $D$ refers to the raindrop diameter in mm. $N_0 = 8000 m^{-3}mm^{-1}$ and $\Lambda = 4.1 R^{-0.21} mm^{-1}$ are distribution parameters, and $R$ is the rain rate in mm/hr. $N(D)$ is the number density of raindrops of diameter $D$ in a unit volume. This distribution, however, does not align well with experimental data for drops smaller than 1 mm. Joss, Thams, and Waldvogel proposed a similar form [106], differing in constants for drizzle, widespread, and thunderstorm rain cases [43].

The log-normal distribution offers a better estimate of DSD for small raindrops, as

$$N(D) = \frac{N_t}{\sqrt{2\pi} \ln \sigma D} \exp\left[\frac{-\ln^2(D/\bar{D})}{2\ln^2 \sigma}\right] \qquad (2)$$

providing more flexibility than the exponential distribution [107]. Here $N_t$ is the total number of raindrops, $\sigma$ is the standard geometric deviation, and $\bar{D}$ is the geometric mean diameter.

Ulbrich's gamma distribution improves DSD estimation accuracy [107], especially at high rain rates, as

$$N(D) = N_0 D^\mu \exp(-\Lambda D) \qquad (3)$$

where $\Lambda$ is the distribution parameter, $\mu$ is the shape parameter and $N(D)$ is in $m^{-3}cm^{-1-\mu}$. This distribution accurately describes raindrop sizes across different rain intensities, with parameters shown in Table 3. Another case of gamma distribution was proposed by setting the shape parameter $\mu = 2$ [108, 109], as $N(D) = N_0 D^2 \exp(-\Lambda D)$. The normalized gamma distribution was first defined by Willis by introducing three parameters $\mu$, $N_\omega$ and $D_m$ [110] and later presented by Montopoli [111] as

$$N(D) = N_\omega \cdot \frac{6}{4^4} \cdot \left[\frac{(4+\mu)^{(4+\mu)}}{\Gamma(4+\mu)}\right] \cdot \left(\frac{D}{D_m}\right)^\mu \cdot \exp\left[-(4+\mu) \cdot \frac{D}{D_m}\right] \qquad (4)$$

It effectively fits various DSDs and allows analysis of changes in distribution parameters [112].

**Table 3 Parameters for Gamma Distribution [107]**

| Type of Rain | $N_0$ [m$^{-3}$mm$^{-1}$] | [Λ mm$^{-1}$] |
|---|---|---|
| Thawing of Pellets (Hail) | 64500R$^{-0.5}$ | 5.7R$^{-0.27}$ |
| Thawing of Granular Snow (Sleet) | 11750R$^{-0.29}$ | 4.1R$^{-0.2}$ |
| Thawing of Non Granular Snow (Sleet) | 2820R$^{-0.18}$ | 3.0R$^{-0.19}$ |

**Theoretical modelling**

Absorption and scattering are the primary causes of attenuation in outdoor line-of-sight (LoS) THz channels. Assuming a plane wave propagates in a direction and interacts with a raindrop, it induces a transmitted field inside the drop and a scattered field [113]. Using Mie scattering theory, we can model the absorption, scattering, and total extinction cross-sections of raindrops [114-116]. The scattering amplitude depends on factors like frequency, size, material, shape of the raindrop, and polarization of the incident wave. Parameter $Q_a$, the absorption extinction cross section, represents the power absorbed by the raindrop, and parameter $Q_s$, the scattering extinction cross section, represents the power scattered in all directions. Parameter $Q_t = Q_a + Q_s$, the total extinction cross section is directly related to attenuation of the transmitted channel, and can be expressed as $Q_t = -(4\pi/k)\text{Im}\left[\hat{e}, f(\hat{K}_1, \hat{K}_2)\right]$, where $\hat{e}$ is a unit vector of the polarization state, $\hat{K}_1$ represents incident direction and $\hat{K}_2$ represent scattering direction. The term $f(\hat{K}_1, \hat{K}_2)$ refers to a matrix function denoting scattering amplitude and the polarization state of the scattered channel. $\hat{K}_1 = \hat{K}_2$ can be considered when there is only forward scattering component. The absorption and scattering can be modelled by several different methods such as Mie scattering, Rayleigh approximation, depending on the frequency of the channel and the shape of the raindrop. The results of Mie's solution leads to expressions for the scattering and extinction efficiencies of the sphere in the form of converging series [117], given by

$$\xi_s(n,\chi) = \frac{2}{\chi^2}\sum_{l=1}^{\infty}(2l+1)\left(|a_l|^2 + |b_l|^2\right) \quad (5)$$

$$\xi_e(n,\chi) = \frac{2}{\chi^2}\sum_{l=1}^{\infty}(2l+1)\text{Re}(a_l + b_l) \quad (6)$$

where $a_l$ and $b_l$ as functions of $n$ and $\chi$, are Mie coefficients. If the particle size is much smaller than the wavelength of the incident channel $|n\chi| \ll 1$, For very small particles, the Mie theory approaches the Rayleigh approximation [103, 118]. The attenuation coefficient $\alpha$ in dB/km can be determined by integrating over all raindrop sizes as

$$\alpha = 3.3429 \int_0^\infty p(r)\xi_e(r)\cdot \pi r^2 dr \qquad (7)$$

where $p(r)$ is the raindrop size distribution. Using the Marshall-Palmer distribution [119], we calculate the spectral attenuation due to rain, as shown in Fig. 7. Without gaseous attenuation, this curve remains stable above 100 GHz, as confirmed in [58]. Power loss from scattering is less than that from absorption due to the water content in rain, with scattering loss remaining constant above 100 GHz, increasing the difference between absorption and scattering losses with frequency. Similar trends in total attenuation, absorption, and scattering are observed under different raindrop size distributions [50], suggesting consistent channel performance in these conditions. The ITU-R provides another model for gaseous absorption based on the MPM93 physical model [53] and predicts rain attenuation using the empirical equation $\alpha_{rain} = kRr^a$, as fitted from measurement data [97].

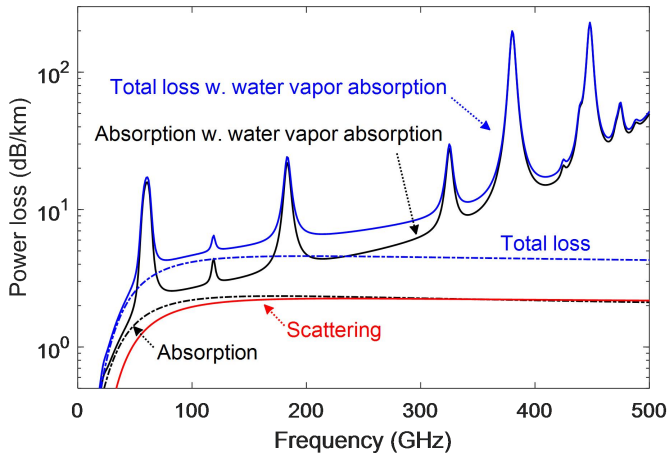

**Figure 7** Attenuation by rain under a rainfall rate of 12 mm/hr (heavy rain) when the Marshall-Palmer distribution employed. ($T = 25^oC$, $P =1013$ hPa and RH = 97%).

**Experimental measurements**

Experimental environments can be categorized into indoor and outdoor settings. To generate stable and controlled rainfall, we constructed an indoor rain chamber in our laboratory [58, 64]. This setup allows us to analyze and observe attenuation caused by raindrops on THz channels. However, the confined space within the rain chamber limits the propagation distance. To mitigate this, mirrors are added to extend the propagation path length, simulating longer distances for studying THz channel performance during rainfall [120]. Outdoor experiments, as reported in [121-128], can evaluate THz channel performance in actual rainfall, modeling it by recording real-time rainfall data during the experiments.

We investigated the effects of indoor rainfall on THz channels [56]. The rain chamber includes a controllable rainfall generator and a path region for channel propagation. Rainfall intensity is controlled by adjusting the air pressure linearly within the chamber, varying the rain rate from 50 mm/hr to 500 mm/hr, with raindrop sizes following a log-normal distribution. The mean raindrop diameter is 1.96 mm with a standard deviation of 0.157. Most raindrops are approximately spherical. The total number of raindrops ($N_{Rr}$) depends on the pressure and rain rate ($R_r$). We designed the measurement setup (see Fig. 6) by utilizing a commercial T-Ray 2000™ THz time-domain spectrometer (THz-TDS) [129-131]. Three gold-plated mirrors reflect the THz radiation multiple times, extending the beam path to about 4 m. To obtain spectral attenuation, the system scans 1000 THz pulses and averages the results.

After applying inverse fast Fourier transformation to the raw data, we observed strong water vapor absorption lines at 0.56 THz, 0.75 THz, and 0.98 THz in both free space and rain conditions, as shown in Fig. 8(a). The phase spectra, illustrated in Fig. 8(b), reveal strong phase jumps corresponding to these absorption lines, with negligible phase shift due to rain except at these frequencies [132].

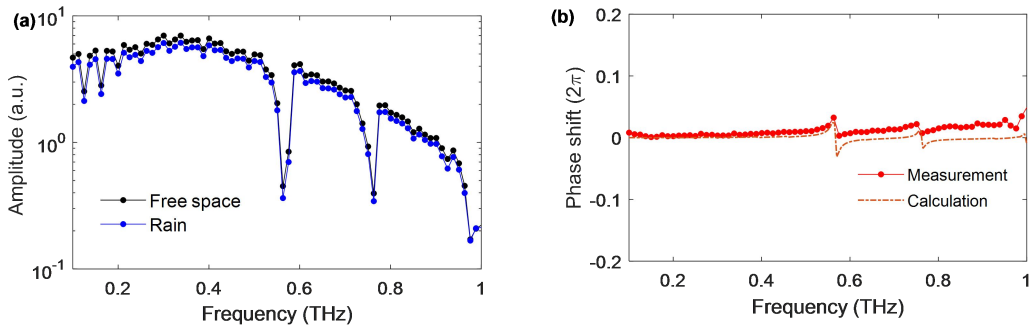

Figure 8 (a) Receive spectra of broadband THz pulses propagating either through free space (red) or rain (blue). (b) Phase spectra of transmitted THz pulses with the same legend as Fig. 8(a). The black curve corresponds to the difference between both after using an unwrapping algorithm.

We also studied the impact of rain on a THz data link using a 16-QAM communication setup with a 5 Gbps data rate, depicted in Fig. 9(a). When the 162 GHz channel propagates through rain at a rate of 350 mm/hr, varying the transmitted power reveals the BER as a function of received power, shown in Fig. 9(b). Below -40.3 dBm, BERs could not be recorded, indicating power loss in the data channel causes BER degradation. The slopes of both curves are almost identical, which indicates that the BER degradation is mainly due to the power loss in the data channel. Fig. 10 compares measured and calculated power loss at 140, 220, 340, and 675 GHz, frequencies within atmospheric transparency windows used for various wireless link setups [133]. Discrepancies between measured data and ITU-R or Mie scattering models with M-P distribution are noted. However, assuming a log-normal distribution for raindrop sizes significantly improves agreement between modeling and experiments, demonstrating Mie scattering's reliability in predicting rain-induced channel attenuation [134].

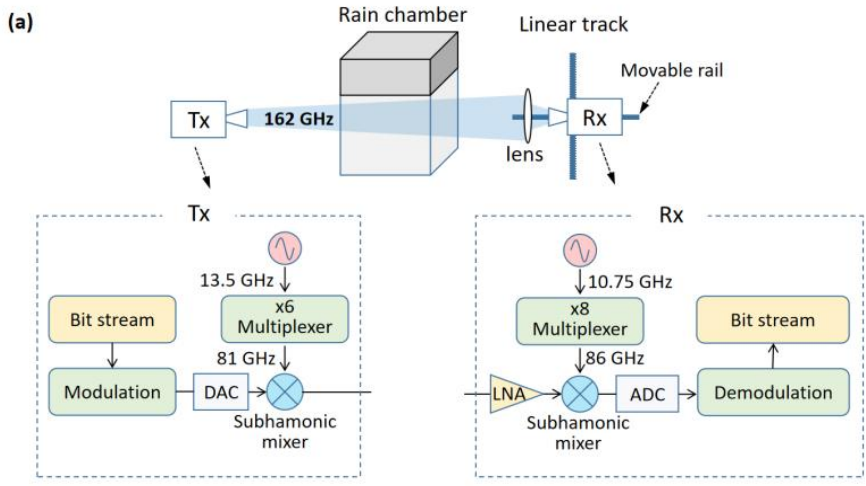

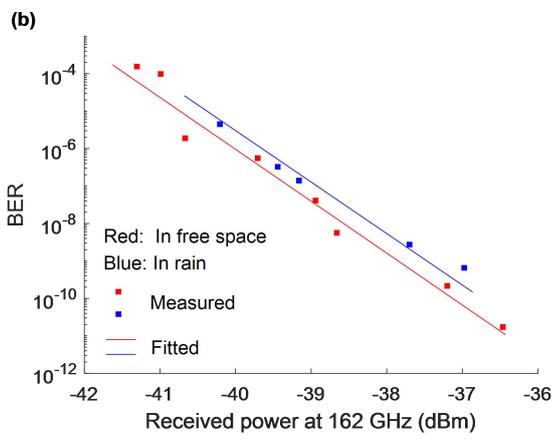

**Figure 9.** Degradation of a THz channel in rain as a function of transmitted power from the antenna at the transmitter side. (a) Schematic of measurement setup. (b) Measured real-time BER performance of the THz link as a function of the received power at a data rate of 5 Gbps. Source: Reprinted from [56] with the permission of Elsevier.

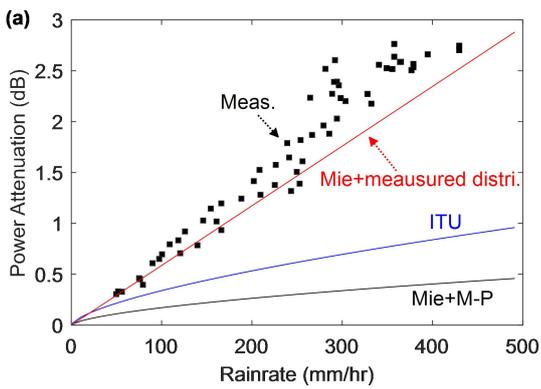
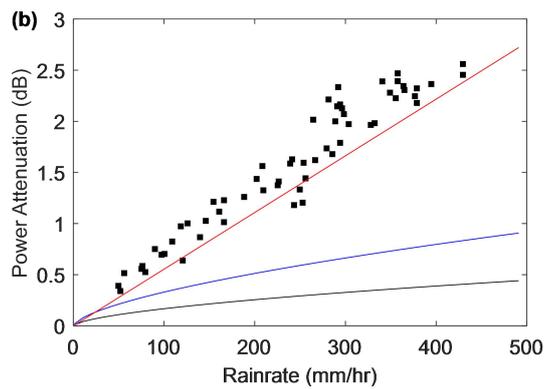

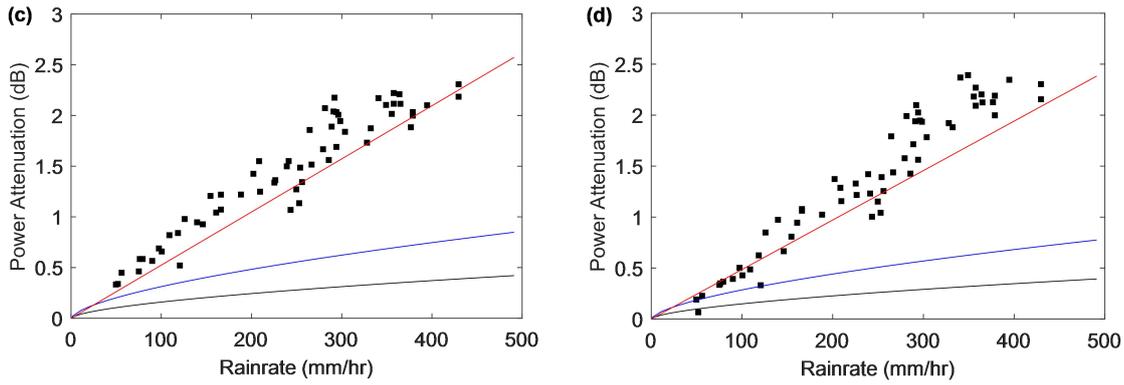

**Figure 10** Power attenuation on wireless channels [56] operating at (a) 140 GHz, (b) 220 GHz, (c) 340 GHz and (d) 675 GHz caused by rain at rain rates between 50 mm/hr and 450 mm/hr over a channel distance of 4m.
**Source: Reprinted from [56] with the permission of Elsevier.**

A key challenge in understanding rain-induced THz channel degradation is the limited range of empirical models, which may not account for factors like spatial structures and climatic variations. To address this, we propose a theoretical model incorporating height-induced DSD variations. Outdoor experiments, conducted at the Beijing Institute of Technology during Typhoon "DuSuri", provided a 41.4-meter-long channel (see Fig. 11). Data was logged at one-minute intervals over 90 minutes, with rain rates recorded every ten minutes. We measured the rain rate distributions, with significant concentrations at 6.8 mm/hr, 7.6 mm/hr, 15.3 mm/hr, and 30.6 mm/hr. We refined the exponential distribution model based on real-time rain rates, corresponding to $N_0$ (mm$^{-1}$m$^{-3}$) values of 1850, 1680, 1000, and 620, and $\Lambda=4R^{-0.21}$ mm$^{-1}$. Using these parameters, we derived the DSD model for rain rate variation and applied Mie scattering theory to ascertain rainfall attenuation. Our computational results aligned well with measured data, validating our DSD correction model. While effective across various rain rates, further optimization is needed for changing conditions.

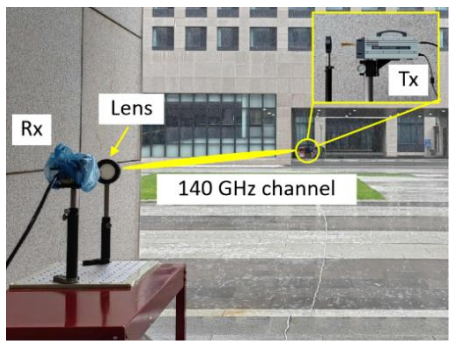

**Figure 11** Schematic representation of the outdoor setup designed to gauge rainfall attenuation experienced by a 140 GHz channel.

**Analysis**

In contrast to outdoor experiments, indoor rain chamber experiments allow for more precise measurement of time-dependent rainfall amounts and enable continuous adjustment of rainfall intensity according to the experimental requirements. This capability ensures that the simulated atmospheric environment can be accurately controlled and adjusted to the desired ideal state. On the other hand, outdoor rainfall simulations provide the most realistic representation of atmospheric rainfall, which is crucial for understanding the real-world impact of rain on THz channels. By utilizing precise and timely measurements along with subsequent data processing, researchers can gain significant insights into how rain affects THz communication in a natural setting. This method also aids in generating more intuitive and effective experimental conclusions. While both experimental approaches share similar principles, their results may differ due to inherent variations in their setups. Nonetheless, both methods have unique advantages that merit further exploration.

Moreover, Mie scattering theory remains the most appropriate theoretical model for these calculations. This model allows flexibility in altering the Drop Size Distribution (DSD) based on the current rain conditions, thus enhancing the accuracy and relevance of the experiments under varying atmospheric scenarios.

**3.2 Snow**

The scattering process in snow requires detailed information on snow parameters like shape, dielectric constant, size distribution, snowfall rate, and temperature [135, 136]. Snow particles often have complex shapes, but generally, their scattering and absorption efficiencies are weakly dependent on shape [137]. Photographic measurements show that the ratio of the maximum horizontal dimension to the height of falling snow particles is near unity, justifying a spherical approximation for simplifying computations [138]. As a mix of ice, air, and water., the dielectric constant of snow relates to the constants and volume fractions of these components. Dry snow, a mixture of ice and air, is described by an empirical formula [139], while wet snow, containing ice, air, and free water, is modeled using a two-phase Polder-Van Santen model [140] or a modified Debye-like model [141] for frequencies above 15 GHz. The dielectric constant of water is calculated using the Double-Debye dielectric model (D3M) [142-145], and pure ice is modeled using a single Debye model [142, 143, 146-148].

Another extremely important parameter in calculating scattering phenomena is the snow size distribution. This can be affected by various microphysical and dynamic processes inside and below cloud layers. In practical applications, empirical mathematical formulas derived from the observed size spectra have been used to approximate natural snow size distributions. Unlike raindrops which follow exponential [105, 144], Gamma [145] or Log-normal distribution [149], snow size distribution is often described by a negative exponential function. The Marshall-Palmer (M-P) distribution function [105] was originally proposed for modeling raindrop size distribution, but has also found applicability for snow particles [150]. Based on results in [150, 151] and parameters in [105], a modified exponential function was developed by Scott [152] with actual snow particle size $r_m$ used. The first negative exponential distribution function was reported by Gunn and Marshall (G-M) based on ground observations of snow [153] and an assessment method used for raindrop size distribution in [105]. Sekhon and Srivastava (S-S) demonstrated an updating [154] by analyzing the data set in [153] with additional snowflake size distribution measurements.

**Theoretical modelling**

Attenuation of THz channels by snow in the near-surface atmosphere is complex due to variations in snow characteristics [155]. Two empirical models predict power attenuation by snow: one for dry snow with a specific attenuation $\alpha_{snow}$ at 0°C as $\alpha_{snow} = 0.00349R^{1.6}/\lambda^4 + 0.00224 R/\lambda$ [156], where $\lambda$ is the wavelength in centimeters; and a second model for both dry and wet snow with $\alpha_{snow} = a \cdot R^b$ (dB/km), which is valid in all kinds of snow conditions and the parameters $a$ and $b$ are different for dry and wet snow [157]. The scattering effects of snow particles on THz channels are analyzed using Mie theory, suitable due to the particle size range of mm to cm [158], which matches or exceeds THz wavelengths. This method describes scattering mechanisms using absorption and scattering coefficients, assuming negligible multiple scattering and independent scattering events. Attenuation in THz channels through snow can be calculated by Eq. (8) using a basic snow size distribution model, the negative exponential distribution function $N(r) = N_0 \exp(-\Lambda r)$, with $r$ being the radius of melted snow particles and should be used in all the formulas except the Scott function (Table 4), where a conversion to equivalent drop size $r_m$ is required [159]. Parameters $N_0$ and $\Lambda$ are two characteristic parameters which can be retrieved by snowfall rate. Different researchers have treated $N_0$ and $\Lambda$, derived from snowfall rates, vary among researchers. Table 4 lists parameters for four size distributions. Snowfall rate is preferred over visibility for indicating snowfall intensity due to variability in snow type and visibility conditions [160].

**Table 4 List of negative exponential snow size distributions.** $Rr$ is snowfall rate in [mm·hr$^{-1}$].

| Distribution | $N_0$ [m$^{-3}$·mm$^{-1}$] | $\Lambda$ [mm$^{-1}$] |
|---|---|---|
| Marshall-Palmer | $16 \times 10^3$ | $8.2Rr^{-0.21}$ |
| Scott | $100 \times 10^3$ | $5.76Rr^{-0.31}$ |
| Gunn-Marshall | $7.6 \times 10^3 R^{-0.87}$ | $5.1Rr^{-0.48}$ |
| Sekhon-Srivastava | $5.0 \times 10^3 R^{-0.94}$ | $4.58Rr^{-0.45}$ |

The power loss due to dry and wet snow for a typical snowfall rate using the Gunn-Marshall snow size distribution is calculated and shown in Fig. 12. Dry snow, treated as a mixture of ice and air, shows minimal absorption at lower frequencies due to its low dielectric constant [139]. The total power loss caused by dry snow is primarily due to scattering, while wet snow's higher water content results in significant absorption and scattering losses. Below 800 GHz, wet snow causes higher total attenuation than dry snow. A comparison between calculated and measured channel attenuation at 300 GHz for dry and wet snow is conducted [50], using the Gunn-Marshall size distribution[50]. Discrepancies are attributed to humidity and turbulence effects, which are not included in the calculations. Thus, we did the calculation on total power loss by dry and wet snow with water vapor absorption (WVA) included. As frequency increases, gaseous absorption significantly raises total power loss, while scattering loss becomes less dominant. Wet snow, modeled using a Debye-like approach, shows higher absorption than scattering due to its water content, making absorption the main contributor to total power loss above 100 GHz [141]. This means the scattered power by snow would not be frequency-dependent in this frequency range.

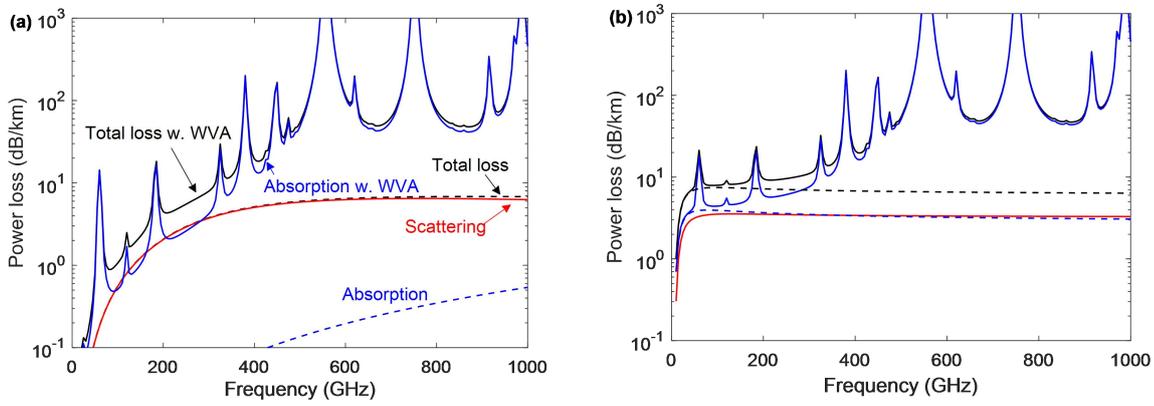

**Figure 12.** Power loss due to (a) dry snow at -1ºC and (b) wet snow (water content 25%) at 0ºC, under the G-M distributions with a snowfall rate (equivalent rainfall rate) of 10 mm/hr. (*P*=1013hPa, RH=97%). (b) keeps the same legend with (a).

Theoretical models of channel power attenuation help understand THz channel performance. However, bit-error-ratio (BER) analysis is crucial for evaluating link reliability, influencing antenna design, signal processing, and error correction. We examined BER performance under various liquid water equivalent (LWE) precipitation rates. Both ITU and Scott models show increased consistency at higher frequencies, with Scott more accurately predicting power loss [40]. Thus, we employed the Scott model and analyzed BER under clear and snowy conditions using ASK and 16-QAM modulation at 140 GHz, 220 GHz, by using established mathematical expressions, $\text{BER}_{ASK} = Q\left(\sqrt{2 \cdot SNR}\right)$ and $\text{BER}_{16-QAM} = 4/\sqrt{M} \cdot Q\left(\sqrt{(3 \cdot SNR)/(M-1)}\right)$, incorporating the *Q*-function to denote the probability of a Gaussian random variable exceeding a specified value, with modulation order denoted by *M*. As shown in Fig. 13, lower frequencies and simpler modulation schemes (e.g., ASK) showed better BER performance due to less power loss. Under certain conditions, error-free transmission (BER<10$^{-10}$) is achievable at lower frequencies with simpler modulation, emphasizing strategic frequency and modulation choices for THz communications in adverse weather. Besides, optimal THz performance occurs with LWE rates below 1 mm/hr, typical of light snowfall, minimally impacting communication over 1 km. Reliability decreases with increased distance or snowfall intensity. Forward error correction (FEC) can enhance resilience within certain thresholds, but its effectiveness depends on coding complexity and redundancy.

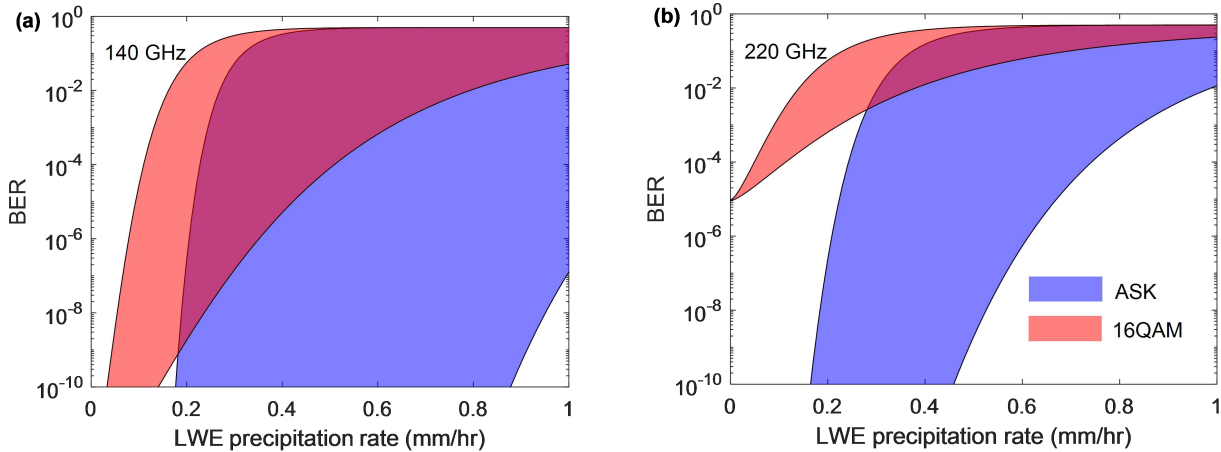

**Figure 13** Predicted average BER performance of the channels operating at (a) 140 GHz and (b) 220 GHz. The upper and lower bounds of the predicting area correspond to the predictions by ITU model (dry) and Scott model, respectively. Channel distance 1 km, relative humidity RH 50 %, temperature 0°C, transmitted power 20 dBm, noise level of receiver -60 dBm; the gain at the transmitter and receiver side are identical and equals to 40 dB (combination of antenna and lens).

**Experimental measurements**

Conducting snow chamber experiments is challenging due to the need for sub-zero temperatures and control over snowflake shapes and sizes. Therefore, we conducted outdoor measurements to evaluate THz channel performance in real snowfall. On 13 March 2018, during a significant snowstorm in New England [161], we set up an 11-meter outdoor LoS THz data link with a 200 GHz carrier wave modulated using ASK at 1 Gb/sec at Brown University [162]. The first 3 meters of the beam path were protected under an overhanging roof, while the remaining 8 meters were exposed to snowfall (Fig. 14a). We measured received power and BER before and during the snowfall. Snowfall reduced received power by approximately 2 dB, requiring higher power to maintain the same error rate (Fig. 14b). Using our measurements of snowfall rate (3.5 mm/hr) and snow density (0.52 g/cm$^3$), we estimated that 27 snowflakes pass through the beam per second, each taking about 32 ms to traverse the beam diameter of 5 cm. Wet snow, modeled using Debye's mixture theory [163, 164], showed a predicted worst-case attenuation of about 3 dB at 200 GHz, closely matching our measured attenuation of 1.8±0.5 dB [161].

We then calculated the degradation of the received power and the BER due to snow. We use the black data points from Fig. 14 (no snow) as a reference to compute the change in the BER due to power loss. We assume that $\Delta t = 0.032$ s is required for all 27 snowflakes to pass through the beam simultaneously, while the beam remains clear for the remainder of the integration time of the BER tester (1 sec) or the power meter (40 ms). Then, the average values for the measured power and BER are given. For the bit error rate, $BER_S$ is calculated from the reference data (black line in Fig. 14(b)) based on a signal loss of 2.95 dB. We predict an average attenuation of 2.2 dB, which is in good agreement with the measured value of 1.8±0.5 dB. The predicted BER result, shown as the red curve, matches very closely to the experimental result.

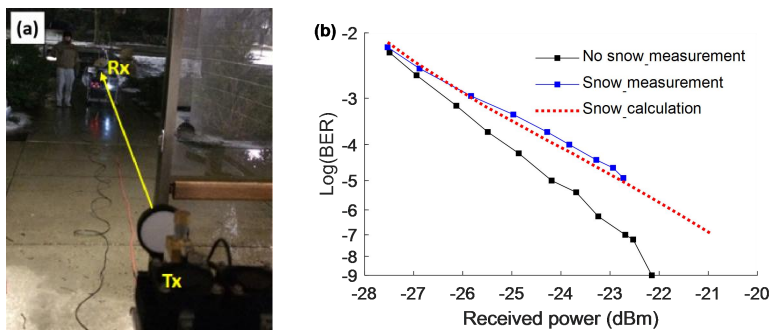

Figure 14 (a) THz wireless link measurement setup; (b) Measurement with (blue) and without snow (black), and the calculated result (red). Source: Reprinted from [161] with the permission of Springer Nature.

We also analyzed THz channel performance in snowy environments on the rooftop of Building 4 at Beijing Institute of Technology (BIT). Using a fixed point-to-point setup, we measured 220 GHz channels. The components were positioned at a range of 11 meters due to lower transmit power and increased path loss at higher frequencies. Our equipment setup, including a Ceyear 1465D signal generator and horn antennas with dielectric lenses, allowed for precise data collection at 7 Hz (see Fig. 15(a)). Measurements were taken on December 13, 2023, at 0°C and 60% RH, indicative of wet snow. Using the cumulative distribution function (CDF) of received power, we observed a power reduction trend correlating with increasing LWE (liquid water equivalent) rates, highlighting the need for counteracting methodologies in THz communication systems. Our CDF analysis showed minimal scattering effects from snowfall, with Rician distribution fitting both clear and snowy conditions. High K factors in snowy conditions indicated dominant LoS components, suggesting minimal multipath effects, consistent with prior studies [165, 166].

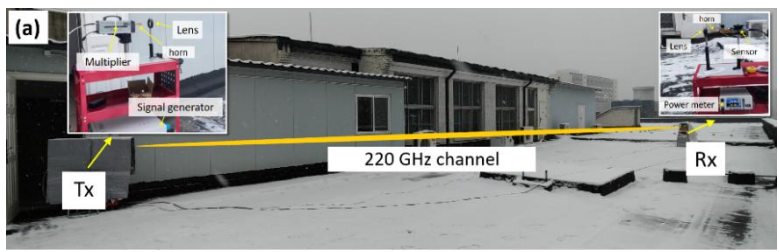

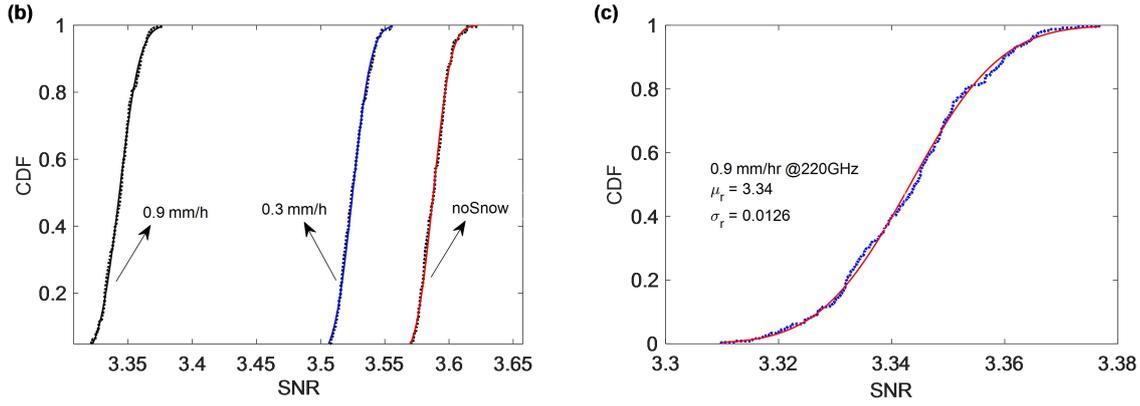

**Figure 15** THz channel measurement setup implemented in the campus of Beijing Institute of Technology (BIT). (a) Outdoor channel on the rooftop of Building 4 at BIT with both transmitter (Tx) and receiver (Rx) safeguarded by waterproof coverings; CDF profile for received SNR with and without snowfall with operating frequencies at (b) 220 GHz, alongside the fitted CDF to the measured data at (c) 220 GHz in snowfall conditions.

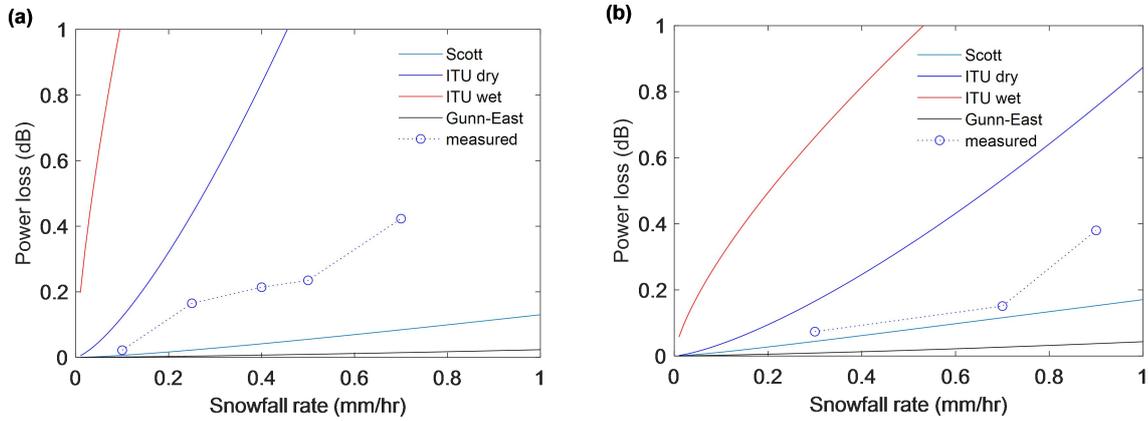

**Figure 16** Power loss relative to the LWE precipitation rate for the 120 GHz and 220 GHz channels, over distances of 21 m and 11 m, respectively. (b) keeps an identical legend and marks as (a).

To quantify power loss due to snow, we monitored temporal variations and aggregated data to derive average values. We employed ITU-R P.1817-1 [167], Gunn-East [156], and Mie scattering theories with the Scott distribution for snowflakes. The Scott distribution, using a negative exponential function [152], provided refined analysis compared to other distributions like Marshall-Palmer and Gunn-Marshall. The Mie scattering theory, with Debye's mixture theory for dielectric properties, and ITU-R models for gaseous absorption, offered nuanced understanding (Fig. 16). Our theoretical predictions revealed that while the ITU model overestimates power loss and the Gunn-East model underestimates it, the Mie scattering theory provides a closer approximation, albeit with occasional underestimations as evidenced in previous findings [50]. Empirical data fell between the Scott and ITU model predictions, suggesting a combined approach could improve accuracy. Wet snow parameters were considered due to the high humidity and near-freezing temperatures during measurements. The ITU model for dry snow closely matched our data, prompting its inclusion in further analysis. Estimated snowflake wetness was assumed to be 10% (Fig. 16).

**Analysis**

Snow particles present a complex challenge for THz communications due to their varied shapes, sizes, and compositions. Modeling snow attenuation typically involves approximating snowflakes as spherical particles composed of mixtures of ice, air, and water. The dielectric properties and size distributions of snow particles are critical factors in determining attenuation. Simulated snow environments are difficult to achieve in laboratory settings due to several factors. These include the need for sub-zero temperatures, precise control over snowflake particle shape and size, and the challenge of replicating the dynamic nature of falling snow. Additionally, maintaining consistent atmospheric conditions such as humidity and temperature while generating artificial snow is extremely challenging. As a result, most related research relies on outdoor measurements in natural snowfall conditions.

However, outdoor measurements come with their own set of difficulties. These include unpredictable weather patterns, variability in snowfall intensity and composition, and the challenge of maintaining consistent experimental conditions over time. Environmental factors such as wind speed, temperature fluctuations, and changes in humidity can all impact measurements. Furthermore, the need for specialized equipment that can withstand harsh winter conditions adds another layer of complexity. Precise quantification of snow effects on THz channels remains difficult due to the complex and variable nature of snow itself. Factors such as snowflake size distribution, water content, and the dynamic behavior of falling snow all contribute to this complexity.

Due to these challenges in outdoor measurements, there are some discrepancies between different publications in their findings on THz channel degradation in snow. Norouzian et al. [168] reported an attenuation of 10 dB/km at 300 GHz with a snowfall rate of 5 mm/hr, while Sen et al. observed much higher attenuation, up to 13.34 dB, for a 140 GHz channel over just 70m at a lower precipitation rate of 0.45 mm/hr [166] and our own study yielded different results still, with attenuation around 2 dB at 200 GHz and a snowfall rate of 3.5 mm/hr over an 8m channel distance [161]. These disparities highlight the complex nature of snow's impact on THz wireless channels. While general trends such as increased attenuation with snowfall rate and frequency are consistently observed, the exact quantification of these effects varies between studies. This variability highlights the complex nature of snow's impact on THz wireless channels. Evidently, there is still a need for more comprehensive studies and standardized measurement techniques to fully characterize the impact of snow on THz wireless communications.

### 3.3 Atmospheric turbulence

In clear weather, where attenuation and scattering by airborne particulates are minimal, THz channel propagation can still be impacted by atmospheric turbulence. Solar radiation heats the Earth's surface, causing the air near the ground to become warmer and lighter than at higher altitudes. This temperature difference leads to thorough mixing of warmer and cooler air, resulting in random temperature fluctuations [169, 170]. Atmospheric turbulence is caused by these temperature inhomogeneities, which can be visualized as air pockets of varying sizes and temperatures. These pockets act like prisms with different refractive indices [89, 94], altering the channel beam's direction as it propagates through the atmosphere, leading to fluctuations in received channel power and phase front distortion [171-173]. Atmospheric turbulence can be characterized by the inhomogeneity of the refractive index along the beam path due to spatial and temporal temperature fluctuations. Scintillation occurs when the propagating channel beam suffers intensity fluctuations at the receiver side, especially over long paths (> 1 km) parallel or near to the ground, leading to time-varying fading effects on the order of milliseconds. The effects of atmospheric turbulence include [174, 175]:

- Scintillation: Distortion of the channel beam's phase front due to refractive index variations, causing irradiation fluctuations.

- Beam Spreading: Diffraction in the propagation path spreads the channel beam.

- Beam Wandering: Deflection of the laser beam when turbulence eddies are larger than the beam diameter.

- Beam Steering: Angular deviation causes the beam to move out of the receiver's aperture range.

Atmospheric turbulence can be described based on turbulence and laminar flow theories. When the Reynolds number exceeds a critical value, turbulent flow occurs [176]. The Navier-Stokes equations can describe turbulence, but their nonlinear nature makes them complex. Kolmogorov theory simplifies this by using energy cascade theory, explaining turbulence through dimensional analysis and approximations [94]. When wind velocity exceeds a critical point, air pockets of different temperatures and sizes form and distribute randomly. Large air pockets break into smaller ones due to inertial forces, with wind shear or convection forming large pockets, and small pockets dissipating energy. These small air pockets act like prisms, causing random interference effects and wavefront distortion (see Fig. 5). These air pockets are statistically uniform and isotropic, smaller than the outer scale $L_0$, which increases linearly with respect to the height above the ground to about 100 meters. This implies a constant mean velocity field, with correlations between random fluctuations depending only on separation, not observation points. In Kolmogorov theory, the longitudinal structure function of wind velocity along the propagation path follows a power law equation [177, 178] as

$$D_v(L_{ij}) = \left\langle (v_i - v_j)^2 \right\rangle = \begin{cases} C_v^2 L_{ij}^{2/3}, & l_0 \ll L_{ij} \ll L_0 \\ C_v^2 l_0^{-4/3} L_{ij}^2, & 0 \leq L_{ij} \ll l_0 \end{cases} \quad (8)$$

Here the velocity structure constant $C_v^2$ measures turbulence energy, with $v_i$ and $v_j$ being the velocity components at points *i* and *j*. Turbulence can also be characterized using the temperature structure function, following a similar power law equation [94, 179]

$$D_T(L_{ij}) = \langle (T_i - T_j)^2 \rangle = \begin{cases} C_T^2 L_{ij}^{2/3}, & l_0 \ll L_{ij} \ll L_0 \\ C_T^2 l_0^{-4/3} L_{ij}^2, & 0 \leq L_{ij} \ll l_0 \end{cases} \quad (9)$$

Here $C_T^2$ is the temperature structure constant, with $T_i$ and $T_j$ representing the temperature at points *i* and *j*. This 2/3 power law behavior corresponds to the inertial subrange, first suggested by Kolmogorov [94]. At small distances, the quadratic behavior of the structure function can be derived using a Taylor series. The temperature structure function, based on the temperature gradient along the laser beam path, is a convenient and accurate method for characterizing atmospheric turbulence in our research.

There have been many statistical models characterize atmospheric turbulence and intensity fluctuations, including log-normal, gamma-gamma, K-distribution, and negative exponential models. But none of them can be suitable for all conditions due to turbulence's non-stationary nature [94]. The log-normal model, defined by a signal parameter related to weather measurements, is widely used for weak turbulence due to its simplicity and good experimental fitting. The gamma-gamma model addresses both strong and weak turbulence [179-181], while the K-distribution suits strong turbulence [182-184]. The negative exponential model is used for saturation regimes [185].

The log-amplitude of channel intensity follows a Gaussian distribution under the central limit theorem [184]. The intensity *I* follows a log-normal distribution as

$$p(I) = \frac{1}{\sqrt{2\pi\sigma_l^2} I} \exp\left(-\frac{\left(\ln\left(\frac{I}{I_0}\right) - E[l]\right)^2}{2\sigma_l^2}\right). \quad (10)$$

Here, $\sigma_l^2$ is the log-intensity variance, and $E[l]$ is the mean log-intensity [186]. The scintillation index, the variance of irradiance fluctuation scaled by the mean irradiance squared, under Rytov approximation, is

$$\sigma_R^2 = 1.23 C_n^2 k^{7/6} L^{11/6} \quad (11)$$

with weak fluctuations corresponding to $\sigma_R^2 < 1$, and strong fluctuations associating with $\sigma_R^2 \gg 1$. It is considered as a representation of turbulence strength related to $C_n^2$ and the path length *L* [94, 177, 187].

The gamma-gamma model attributes channel intensity fluctuation to small and large scale atmospheric effects [188]. Small scales cause scattering (scintillation) by air pockets smaller than the Fresnel zone $R_F = (L/k)^{1/2}$ or the coherence radius $\rho_0$, while large scales lead to refraction by larger air pockets. The gamma-gamma distribution function for received channel intensity can be expressed as

$$p(I) = \int_0^\infty p(I|I_x) p(I_x) dI_x$$
$$= \frac{2(\alpha\beta)^{(\alpha+\beta)/2}}{\Gamma(\alpha)\Gamma(\beta)} I^{\frac{\alpha+\beta}{2}-1} K_{\alpha-\beta}\left(2\sqrt{\alpha\beta I}\right), \quad I > 0 \quad (12)$$

with $\Gamma(\cdot)$ standing for the gamma function and $K_{\alpha-\beta}(\cdot)$ representing the modified Bessel function of the 2nd order of $\alpha - \beta$. The $\alpha$ and $\beta$ are effective number which differs for plane and spherical waves [189-192]. The gamma-gamma turbulence model can be used to characterize turbulence with any strength from weak to strong. While, the K-distributed model, as a special case of gamma-gamma model, can only characterize strong turbulence [193], when the scintillation index ranges from 2 to 3 or the propagation length is larger than 1 km [194]. The distribution function can be expressed by setting $\beta = 1$, as

$$f_K(I) = \sum_{p=0}^{\infty} \left[ a_p(\alpha,1) I^p + a_p(1,\alpha) I^{p+\alpha+1} \right] \quad I > 0 \quad (13)$$

where, the effective number $\alpha$ lies between 1 and 2 [195]. The negative exponential model is used for intensity fluctuation in saturation regimes [179, 196], as

$$f_{NE}(I) = \exp(-I), \quad I > 0 \quad (14)$$

when the mean intensity $E[I] = I_0$ is assumed to be unity in the saturation regime. We observed that, the mean values of the distribution reduce significantly with the intensity as $I_0$ decreases.

**Theoretical modelling**

Atmospheric refractive index is significantly influenced by temperature, humidity, and pressure, with temperature having the most substantial impact on the refractive index in the optical wave band [197]. This sensitivity to small-scale temperature fluctuations makes the refractive index a critical parameter for channel beam propagation. The refractive index $n$ can be approximated for the THz frequency band as $n_{THz} = 1 + 77.6/T \cdot [P_a + 4810 P_v/T] \times 10^{-6}$ [198], where $T$ is the temperature in Kelvin, $P_a$ is atmospheric pressure in millibars, and $P_v$ is water vapor pressure in millibars. The water vapor pressure can be calculated as $P_v = \rho T/216.7$ with water vapor density $\rho$ in unit of g/m³. Pressure fluctuations are generally negligible, making temperature the primary factor for refractive index fluctuations in clear weather conditions [94, 199, 200]. To characterize atmospheric turbulence, the refractive index structure parameter $C_n^2$ can be employed, varying with altitude $h$ [200], as

$$C_n^2(h) = 0.005(v/27)^2 (10^{-5}h)^{10} \exp(-h/1000) + 2.7 \times 10^{-6} \exp(-h/1500) + \hat{A}\exp(-h/100) \quad (15)$$

where $\hat{A}$ is a normal value of $C_n^2(0)$ at the ground level in m$^{-2/3}$ with wind velocity $v \sim 21$ m/s. $C_n^2$ is nearly constant for horizontally propagating fields, with typical values ranging from $10^{-12}$ m$^{-2/3}$ for strong turbulence to $10^{-17}$ m$^{-2/3}$ for weak turbulence [140]. This parameter measures the strength of refractive index fluctuations, calculated using the temperature structure function $C_n^2 = (dn_1/dT)^2 C_T^2$ [94]. The temperature difference is obtained from point measurement of the mean square temperature difference between two temperature sensors at two different points, as we mentioned above, which allows us to determine the parameter $C_T^2$ for any given length $L$. The $C_n^2$ can be inferred in terms of power spectrum of refractive index fluctuations $\Phi_n(k) = 0.033 C_n^2 k^{-11/3}$ [179, 201], with $2\pi/L_0 \ll k \ll 2\pi/l_0$. It is valid for the inertial sub-range, where $k$ is the spatial wave number. Then, Tatarskii and von Karman further refined this equation for a wider range of $k$ values [202].

Evaluating turbulence is complex due to the nonlinear behavior of atmospheric quantities like temperature, pressure, and wind velocity. Thus, turbulence is often characterized using statistical distributions of received irradiance. Two simplifying assumptions are made [203, 204]: (1) The atmospheric free space communication channel is non-dispersive for wave propagation. (2) The mean energy in the absence or presence of turbulence is the same. Then, there are several models describe atmospheric turbulence attenuation, with the Rytov approximation and Andrew's method being prominent. The Rytov approximation links the refractive index structure parameter to the relative variance of channel intensity. For weak turbulence, the scintillation variance is given by $\sigma_\chi^2 = 23.17 \cdot C_n^2 \cdot k^{7/6} \cdot L^{11/6}$ [205]. So, the attenuation due to scintillation [206] can be expressed as

$$\alpha = 2\sigma_\chi = 2\sqrt{23.17 \cdot C_n^2 \cdot k^{7/6} \cdot L^{11/6}} \quad (16)$$

This method, however, does not account for receiver aperture effects, which are significant for turbulence studies.

Andrew's method, based on detailed mathematical analysis by Larry C. Andrews [207], expresses turbulence attenuation as [147, 149]

$$\alpha = 10 \cdot \log\left|1 - \sqrt{\sigma_I^2(D)}\right| \quad (17)$$

where, $\sigma_I^2$ is the scintillation index [171] and $D$ is the receiver diameter.

Air turbulence and humidity fluctuations can distort the THz beam wavefront, causing scintillations that attenuate beam power and degrade channel performance [157, 208]. Snow particles also induce scintillation, with strength depending on their terminal velocity, measured as 1.0 to 1.5 m/s for dry snow[158] and 5 to 6 m/s for wet snow [138]. The refractive index structure parameter $C_n^2$ quantitatively measures turbulence strength. Fig. 17 shows that the scintillation loss increases exponentially with $C_n^2$, exceeding 0.5 dB/km for strong turbulence ($C_n^2 > 10^{-13}$). This highlights the need to consider scintillation loss in calculations, especially over long distances up to several kilometers, which has been achieved in several publications [60, 209].

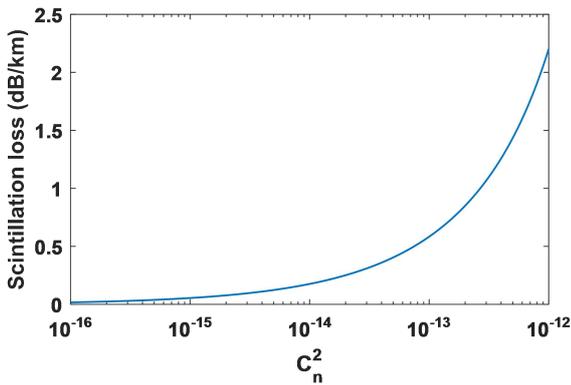

**Figure 17 Scintillation attenuation for THz wave under different turbulence strengths.**
**Source: Reprinted from [50] with the permission of Springer Nature.**

**Experimental measurements**

Measuring and analyzing the impact of atmospheric turbulence on THz communication channels under outdoor conditions is challenging due to the long observation times required and the difficulty in replicating similar atmospheric conditions for independent verification. We emulate atmospheric turbulence in a controlled laboratory setting, using a Mach-Zehnder Interferometer (MZI) with visible light co-propagating with the THz and IR beams. Fig. 18(a) shows the MZI setup embedded in our test bed. A He-Ne laser beam (632.8 nm, 2 mm diameter) is split into two parts by beam splitter BS1 and recombined by BS2, forming localized interference fringes. One part co-propagates with the THz and IR beams at a small angle (< 3°), while the other part propagates through ambient air, isolated from the turbulence chamber (90×40×30 cm³). Air flows into the chamber at constant velocity and temperature, monitored by three thermistors positioned a few centimeters below the beam path, without blocking them. These sensors have a thermal response time $\tau$ of approximately 3 seconds, preventing rapid temperature fluctuation tracing. The air temperature can be set to 32°C (cold), 55°C (warm), and 70°C (hot), with adjustable airspeeds of 28.6 m/s (low) and 41.6 m/s (high). This setup generates air turbulence strong enough to detect degradations in the THz channel.

Changes in the refractive index of turbulent air inside the chamber cause phase changes and beam deflection. The total phase change represents the accumulated turbulence impact. To investigate this, we use beam splitter BS3 to split the visible beam and direct parts of it, after magnification, onto two photodetectors (D1 and D2). The detectors, with apertures about 500 μm wide (compared to the ~5 mm interference fringes), measure medium intensity (D1) and peak intensity (D2). Comparing D1 and D2 allows us to analyze phase changes by measuring power variations. Since refractive index changes are similar across all three wavelength bands, visible light measurements estimate cumulative refractive index fluctuations for the THz channels.

Upon introducing air into the turbulence chamber, we observed typical THz channel attenuation evolution (Fig. 18(b)). Ten seconds after starting the recording, high-speed hot air was launched into the beam's path. The THz beam formed a speckle pattern at the receiver, causing a small signal impact (~0.15 dB), detectable above our sensitivity limit. After turning off the air supply at 20 seconds, the transmitted THz power recovered to its original level. A strong correlation between attenuation and recorded BER was evident. The BERs verified that air turbulence has a small impact on the signal, with a clear trend between channel attenuation and BER. Temperature evolution in the chamber, with de-convoluted recordings compensating for the sensor's slow response time [66], showed that attenuation reduced immediately when airflow stopped, but the temperature decayed more slowly. This indicates that channel power attenuation is mainly due to the constant flow of air pockets with different temperatures, achieving thermal equilibrium once airflow stops, without producing a homogeneous temperature before reaching the beams, causing refractive index variations. Besides, using Eq. (17) and a chamber length $L$=90 cm, we compared experimentally and theoretically determined attenuations and a linear correlation was observed, with some divergence likely due to uncertainty in the effective turbulence length.

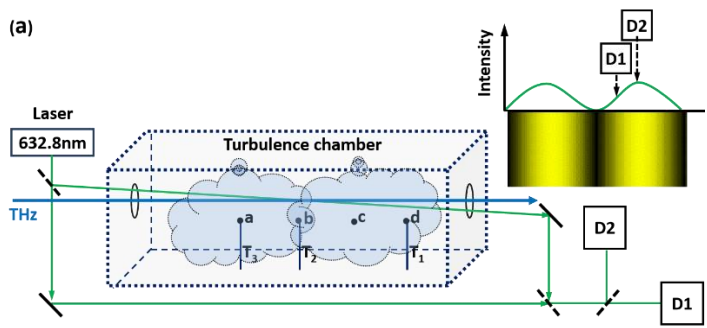

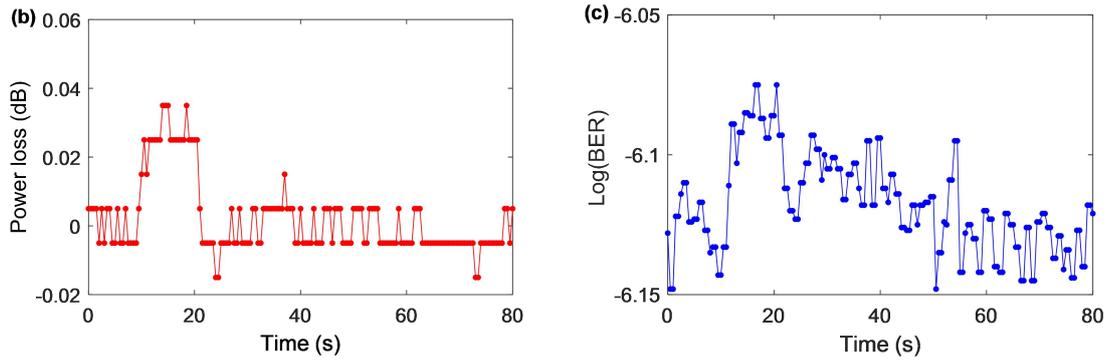

Figure 18 Schematics of a Mach-Zehnder interferometer setup to characterize turbulences with visible light. Inset: Position of photodectors D1, D2 relative to interference fringe intensity. (a) Attenuation of THz channel power as the function of time, for warm air at high speed (b) corresponding Log(BER) of THz channel as the function of time.

**Analysis**

Atmospheric turbulence presents a complex challenge for THz communications due to its dynamic nature and its effects on signal propagation. Even in clear weather conditions, temperature fluctuations in the air can cause variations in the refractive index, leading to phenomena such as scintillation, beam spreading, wandering, and steering. These effects can potentially degrade the performance of terahertz wireless links. While simulating atmospheric turbulence in laboratory settings is possible, it presents its own set of challenges. Creating controlled and reproducible turbulence conditions that accurately mimic real-world scenarios is difficult. Factors such as maintaining consistent temperature gradients, controlling air flow patterns, and scaling the effects to match outdoor conditions over longer distances are all significant hurdles. As a result, many studies rely on a combination of laboratory experiments with limited turbulence chambers and theoretical modeling.

In outdoor measurements, the unpredictable and constantly changing nature of atmospheric conditions makes it challenging to isolate and quantify turbulence effects. Variations in temperature, humidity, wind speed, and other atmospheric parameters can all influence turbulence strength and characteristics. Additionally, these effects can vary significantly over time and distance, making long-term or long-range measurements particularly difficult.

Due to these challenges in both simulated and real-world measurements, there are some discrepancies between different publications in their findings on terahertz attenuation due to turbulence. While general trends, such as increased effects with stronger turbulence and longer propagation distances, are consistently observed, the exact quantification of these effects varies between studies. For long-distance links or in strong turbulence conditions, these effects should still be carefully considered in system design and performance predictions.

### 3.4 Comparative analysis of different weather conditions

The impact of various weather conditions on THz channel propagation varies significantly, each presenting unique challenges and effects. Rain, snow, and atmospheric turbulence are the primary weather phenomena that affect THz communications, and their influences differ in nature, severity, and frequency dependence. When comparing these weather conditions, several key observations emerge. First, the frequency dependence of weather effects varies. While all weather effects generally increase with frequency, rain and snow attenuation tend to increase more rapidly compared to turbulence effects. This suggests that frequency selection in THz systems must carefully consider the prevalent weather conditions in the deployment area.

The complexity of modeling these weather effects also differs significantly. Rain is relatively easier to model due to more consistent droplet shapes and well-established size distribution models. Snow presents greater modeling challenges due to the complex and variable shapes of snowflakes, as well as the significant differences between dry and wet snow. Turbulence modeling is particularly complex due to its dynamic and non-linear nature, often requiring sophisticated statistical approaches. Measurement challenges also vary across these weather conditions. Controlled experiments are easiest to conduct for rain, more difficult for turbulence, and most challenging for snow due to temperature requirements and particle variability. This disparity in experimental ease has led to an imbalance in the available empirical data, with rain effects being the most thoroughly studied and snow effects the least.

In terms of impact severity, wet snow tends to cause the most severe attenuation, followed by heavy rain, as shown in Fig. 19. Dry snow and turbulence typically have less severe impacts on signal strength but can significantly affect signal quality and stability. However, the relative severity can vary depending on the specific conditions and the frequency of operation. It's important to note that real-world scenarios often involve combinations of these weather conditions. For instance, a storm might involve both rain and strong turbulence, or a winter storm could bring a mix of snow types along with turbulent winds. Therefore, future research should focus on developing comprehensive models that can account for multiple weather effects simultaneously. Such models would be invaluable in designing robust THz communication systems capable of operating reliably in diverse and dynamic atmospheric conditions.

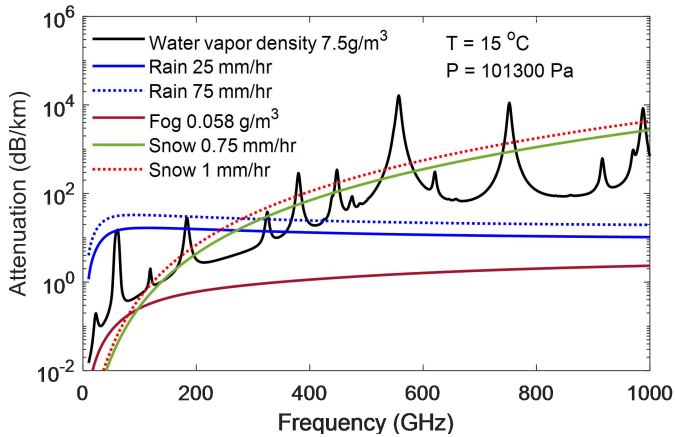
**Figure 19 Atmospheric attenuation spectrum for different humidity levels**

## 4. Eavesdropping risks in adverse weather conditions

The use of THz frequencies, characterized by high-frequency, narrow-angle broadcasts, presents significant challenges for eavesdroppers attempting to intercept or jam signals compared to lower frequency wide-area broadcasts [210, 211]. The narrow beam of THz channels reduces the likelihood of interception, while the high frequency complicates jamming attempts. Despite this, the full scope of THz eavesdropping remains uncharacterized, although it is broadly accepted that high-frequency wireless data links enhance security [43, 212, 213]. Studies at lower frequencies suggest that eavesdroppers must place their antennas within the broadcast sector of the transmitting antenna, implying high directionality makes eavesdropping nearly impossible [212]. In indoor experiments, we demonstrated that an eavesdropper (Eve) could scatter THz signals using a smaller passive object within the beam, redirecting part of the signal to Eve's receiver [214]. This challenges the assumption that Eve needs a large, bulky receiver directly in the beam path. Furthermore, transmitted THz channels in atmospheric environments can be scattered by rain, snow, and turbulence, allowing eavesdropping from positions alongside the beam axis [215].

### 4.1 Rain

We consider a point-to-point outdoor THz channel configuration shown in Fig. 20(a). Alice (the transmitter) sends information to Bob (the legitimate receiver) via a LoS channel, experiencing absorption and scattering losses due to rain. An eavesdropper (Eve) aims to intercept the signal through a NLoS channel scattered from the LoS path. The channel distance $d$ is set at 1 km, with Alice and Bob fixed, while Eve can adjust its position and pointing direction for optimal reception.

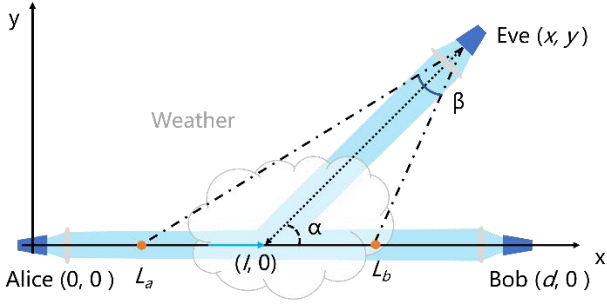

Figure 20 Geographic of the point-to-point THz channel with a security attacker (Eve) in rain or snow, together with the coordinate system of the legitimate (LoS) and eavesdropping (NLoS) channels.

In rain, the LoS channel suffers atmospheric attenuation $G_A$ and divergence attenuation $G_D = 4A/(\pi d^2 \alpha_A^2)$ with $\alpha_A$ as Alice's full divergence angle and $A$ as Bob's effective receiving area. Atmospheric attenuation $\alpha_{atm}$ combines rain attenuation ($\alpha_t$) and gaseous attenuation ($\alpha_g$) and is given by $\alpha_{atm} = \alpha_t + \alpha_g$. The atmospheric attenuation is $G_A = \exp(-\alpha_{atm} d)$. Including divergence attenuation, the total LoS channel gain is

$$G_{\text{LOS}} = G_A G_D = \frac{4A \exp(-\alpha_{atm} d)}{\pi d^2 \alpha_A^2} \qquad (18)$$

For the NLoS channel, a single-scattering model [216, 217] is used. In Fig. 28(b), Alice is at (0, 0), Bob at (d, 0), and Eve at (x, y). The NLoS channel gain $G_{\text{NLOS}}$ [218] can be obtained as

$$G_{\text{NLOS}} = \int_{L_a}^{L_b} \Omega(l) p(\mu) \alpha_{atm} \exp\left\{-\alpha_{atm}[l + \sqrt{(x-l)^2 + y^2}]\right\} dl \qquad (19)$$

with $l$ as the transmission distance before scattering. The bounds $L_a$ and $L_b$ are $L_a = \min\left\{\max\left\{x - \frac{y}{\tan(\alpha - \beta/2)}, 0\right\}, d\right\}$ and $L_b = \min\left\{\max\left\{x - \frac{y}{\tan(\alpha + \beta/2)}, 0\right\}, d\right\}$, respectively, which divides the scattering region. Here, $\alpha$ denotes the angle between the pointing direction of Eve and the positive $x$ axis, and $\beta$ denotes the field-of-view (FOV) full angle of Eve [219]. The solid angle $\Omega(l)$ from the receiving area to the scattering point is $\Omega(l) = \dfrac{A}{\left[(x-l)^2 + y^2\right]^{3/2}} \dfrac{(x-l) + y \tan\alpha}{\sqrt{1 + \tan^2\alpha}}$ and the phase function $p(\mu)$ can be expressed as

$$p(\mu) = \frac{1-g^2}{4\pi}\left[\frac{1}{(1+g^2-2g\mu)^{3/2}} + f\frac{3\mu^2 - 1}{2(1+g^2)^{3/2}}\right],$$ with $\mu = (x-l)/\sqrt{(x-l)^2 + y^2}$ as the cosine of the scattering angle, and $g$ as an asymmetry factor [220].

Secrecy capacity, the maximum data rate ensuring perfect secrecy [221], is

$$C_s = \left[I(X;Y) - I(X;Z)\right]^+ \qquad (20)$$

where, *I(X;Y)* and *I(X;Z)* are the mutual information of LoS and NLoS channels [222], respectively, and can be given by

$$I(X;Y) = q(\lambda_L + \lambda_b)\log(\lambda_L + \lambda_b) + \lambda_b \log(\lambda_b) - (q\lambda_L + \lambda_b)\log(q\lambda_L + \lambda_b) \quad (21)$$

and

$$I(X;Z) = q(\lambda_N + \lambda_b)\log(\lambda_N + \lambda_b) + \lambda_b \log(\lambda_b) - (q\lambda_N + \lambda_b)\log(q\lambda_N + \lambda_b) \quad (22)$$

Here $\lambda_L = \tau\eta G_{\text{LOS}} P/E_p$ and $\lambda_N = \tau\eta G_{\text{NLOS}} P/E_p$ are the mean numbers of detected photoelectrons for LoS and NLoS channels, respectively, with output power $P$, receiver efficiency $\eta$, photon energy $E_p$ and integration time $\tau$. We assume photodetectors are used in this system. They measure the channels by detecting photoelectrons generated by the incident THz radiation. Photodetectors are preferred over thermal detectors for wireless communications as they directly convert photon energy into an electrical signal, providing faster response times. This makes them more suitable for high-speed communications. The SNR of the LoS channel is $\lambda_L/\lambda_b$. We consider outdoor channels at 140, 220, and 340 GHz, collimated with the same beam width. The FOV angle for Eve is 20°, and the receiving areas for Bob and Eve are 1 cm². Atmospheric conditions are set to T=25 °C, P=1013 hPa, and RH=97%. With Eve at (200 m, 10 m), channel gains are calculated (Fig. 21(a)). Solid curves show LoS channel gain $G_{\text{LOS}}$ versus rain rate, decreasing with higher rainfall due to increased attenuation. Dashed lines show NLoS channel gain $G_{\text{NLOS}}$. At 140 GHz, $G_{\text{NLOS}}$ is initially smaller than $G_{\text{LOS}}$, intersecting at 47 mm/hr rain rate, and exceeding $G_{\text{LOS}}$ beyond that due to higher LOS attenuation. For 220 GHz, intersection occurs at 0.5 mm/hr, indicating reduced security due to higher atmospheric attenuation and scattering. At 340 GHz, gains are too small to show.

Secrecy capacity distributions with respect to Eve's position at $Rr$ = 15 mm/hr are shown in Fig. 21(b-d). The color bar indicates safe transmission rate $C_s$ in Gbps, increasing from blue to yellow. For 140 GHz (Fig. 21(b)), dark blue regions ($C_s$ =0) indicate insecure zones. At 220 GHz (Fig. 21(c)), the insecure area shrinks due to higher LoS channel attenuation [223], reducing eavesdropping risk but lowering maximum safe transmission rate to 90 Gbps, compared to 140 Gbps at 140 GHz. At 340 GHz (Fig. 21(d)), the insecure region further reduces, with a maximum safe rate of 60 Gbps.

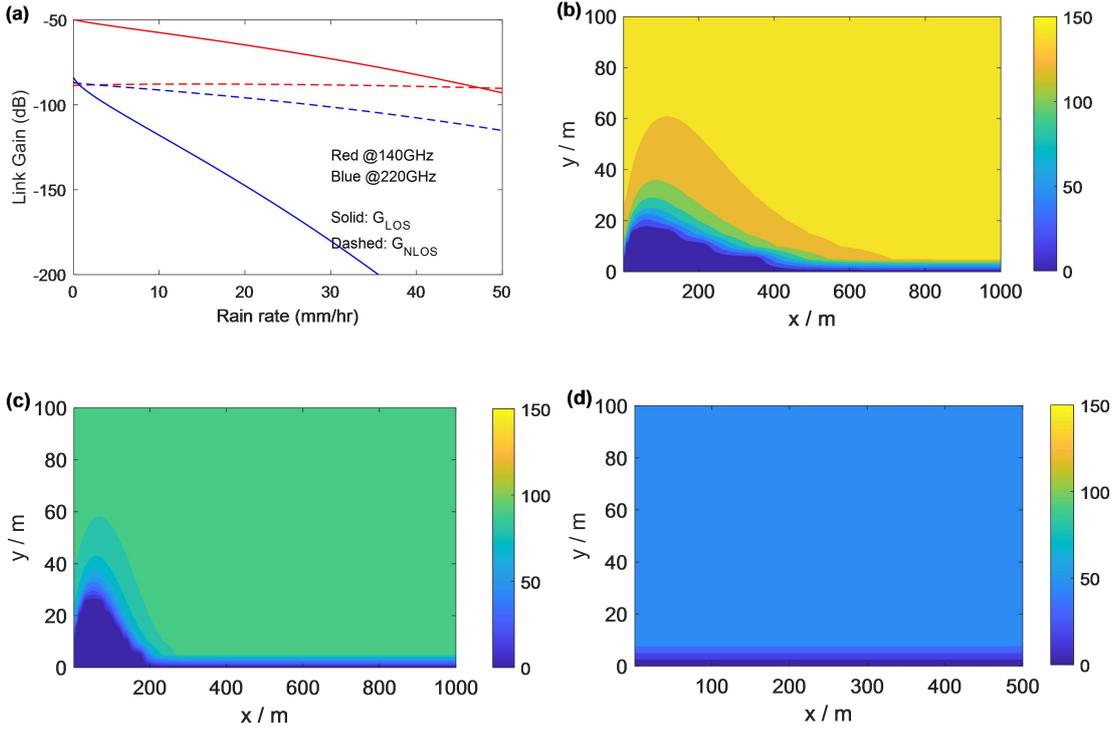

**Figure 21** (a) LoS (solid) and NLoS (dashed) link gain with respect to rainfall rate with carriers at 140 GHz (black), 220 GHz (blue) with Eve at a position of (200m, 10m) and Bob located at (1km, 0m). ($T = 25^oC$, $P = 1013$ hPa and RH = 97%. Secrecy capacity distribution for 2-D positions of Eve when (b) 140 GHz channel, (c) 220 GHz channel, and (d) 340 GHz channel with rainfall rate $Rr = 15$ mm/hr and Bob located at (1km, 0m). ($T = 25^oC$, $P = 1013$ hPa and RH = 97%. The color bar denotes the safe transmission rate in Gbps).

### 4.2 Snow

Using the same parameters as in Fig. 21, we calculate the channel gain for LoS and NLoS channels in dry snow with Eve at (200m, 10m), temperature $T = 0\ °C$, pressure $P = 1013$ hPa, and relative humidity RH = 97%. We use the Gunn-Marshall (G-M) model for snowdrop size distribution. Solid and dashed lines in Fig. 22(a) represent the evolution of $G_{LoS}$ and $G_{NLoS}$, respectively. The gains are higher than in rain due to smaller signal loss in dry snow. Similar to Fig. 21, $G_{LoS}$ decreases with increasing snowfall rate due to higher attenuation. However, there is no intersection between $G_{LoS}$ and $G_{NLoS}$ at 140 GHz, indicating security over 0-50 mm/hr snowfall rates. Intersections appear at $Rr=34$ mm/hr for 220 GHz and $Rr=6.5$ mm/hr for 340 GHz due to higher frequency degradation. Fig. 22(b-d) show the secrecy capacity distributions for Eve's positions in dry snow at $Rr=15$ mm/hr. In Fig. 22(b) for the 140 GHz channel, the insecure region (dark blue) is larger than in Fig. 21(b), despite the maximum safe rate of 140 Gbps, due to higher snow particle density. Equivalent rainfall rate [50] suggests more dry snow particles than rain at the same fall rate, increasing scattering and reducing secrecy. For 220 GHz (Fig. 22(c)), the insecure region grows and maximum safe rate drops to 90 Gbps, attributed to increased scattering, which is confirmed and demonstrated in [224]. At 340 GHz (Fig. 22(d)), the insecure area reduces, but maximum safe rate falls to 60 Gbps due to higher signal absorption reducing Bob's SNR.

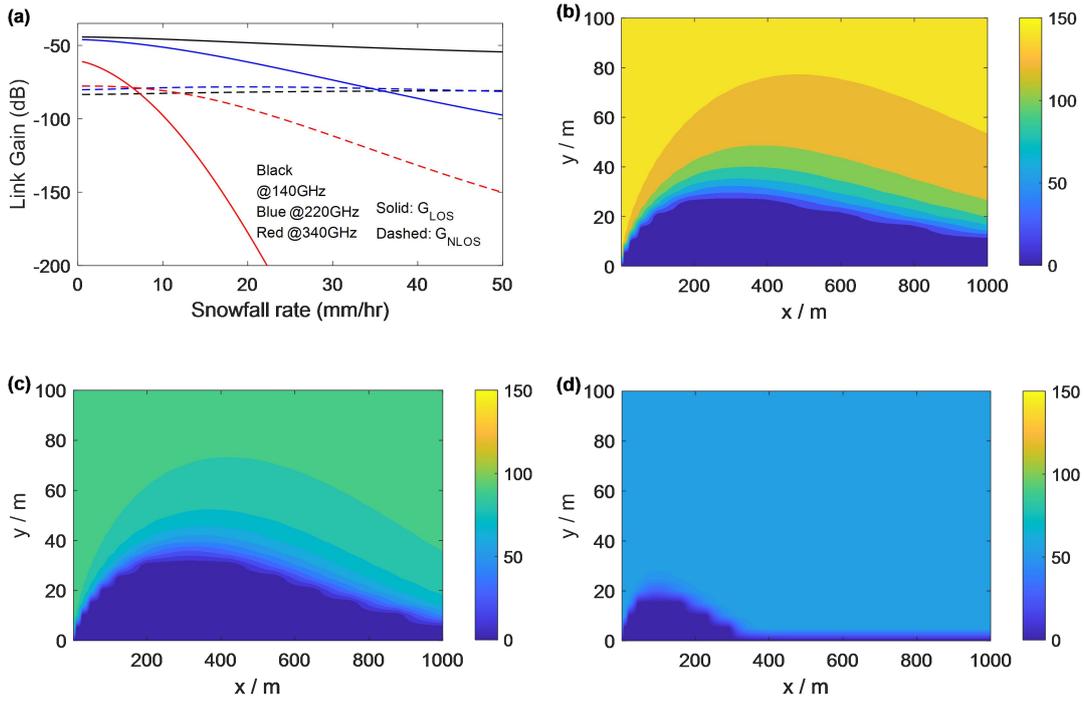

Figure 22 (a) LoS (solid) and NLoS (dashed) channel gain with respect to snowfall rate (equivalent rainfall rate) with carriers at 140 GHz (black), 220 GHz (blue), 340 GHz (red) with Eve at a position of (200m, 10m) and Bob located at (1km, 0m). ($T$ = -1ºC, $P$ =1013 hPa and RH = 97%). Secrecy capacity distribution for 2-D positions of Eve when (b) 140 GHz channel, (c) 220 GHz channel and (d) 340 GHz channel with snowfall rate (equivalent rainfall rate) $Rr$ = 15 mm/hr and Bob located at (1km, 0m). ($T$ = 25ºC, $P$ =1013 hPa and RH = 97%. The color bar denotes the safe transmission rates in Gbps).

In wet snow, with $T = 0 °C$, $P$ = 1013 hPa, and RH=97%, the channel gains for Bob and Eve are shown in Fig. 23(a) using the G-M raindrop size distribution model. For 140 GHz, $G_{LOS}$ is consistently higher than $G_{NLOS}$ up to 50 mm/hr snowfall rate, indicating security. At 220 GHz, an intersection at $Rr$ = 41 mm/hr suggests potential security breaches beyond this rate. At 340 GHz, the intersection occurs at $Rr$ = 24.5 mm/hr, indicating even worse security. At $Rr$ = 15 mm/hr Fig. 23(b-d) shows secrecy capacity distributions for Eve in wet snow. For the 140 GHz link (Fig. 23(b)), the insecure region is smaller than in dry snow (Fig. 23(c)), due to less scattering in wet snow. This trend holds for 220 GHz (Fig. 33(c)) and 340 GHz (Fig. 23(d)), where better link secrecy is observed in wet snow. As carrier frequency increases from 140 GHz to 220 and 340 GHz, the insecure region shrinks, and maximum safe data rate drops from 140 Gbps to 90 Gbps and 50 Gbps, respectively.

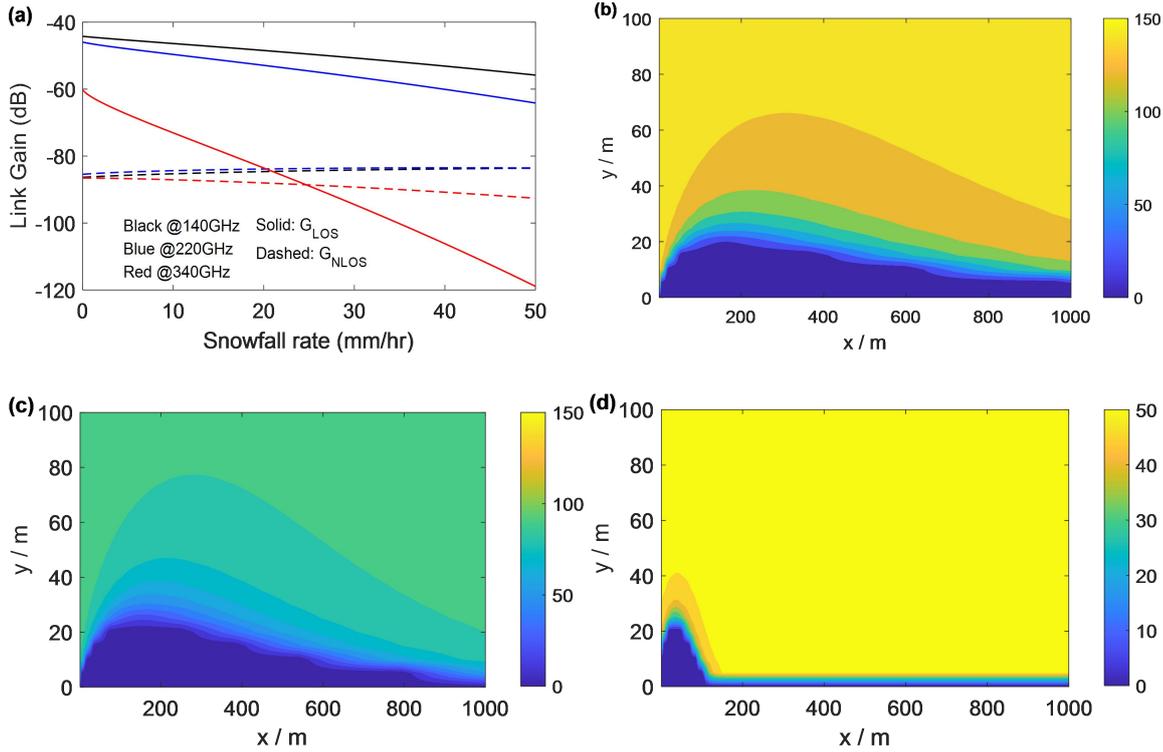

**Figure 23** LoS (solid) and NLoS (dashed) channel gain with respect to snowfall rate (equivalent rainfall rate) with carriers at 140 GHz (black), 220 GHz (blue) and 340 GHz (red) with Eve positioned at (200m, 10m) and Bob located at (1km, 0m). ($T = 0°C$, $P = 1013$ hPa and RH = 97%). Secrecy capacity distribution for 2-D positions of Eve when (b) 140 GHz channel, (c) 220 GHz channel and (d) 340 GHz channel with snowfall rate (equivalent rainfall rate) $Rr = 15$ mm/hr and Bob located at (1km, 0m). ($T = 25°C$, $P = 1013$ hPa and RH = 97%. The color bar denotes the safe transmission rates in Gbps).

### 4.3 Atmospheric turbulence

As noted above, atmospheric turbulence arises from spatial and temporal temperature and pressure inhomogeneities in the air [169, 170]. These can be modeled as numerous air pockets with varying sizes (from a small scale $l_0$ to a large scale $L_0$), temperatures, and pressures, leading to beam divergence. Turbulence-induced signal variation includes a slow component (averaged value $\alpha_t$) due to refractive index variation and a fast component from random refractive index fluctuations[62]. We refer to the channel loss given by Eq. (19) as deterministic attenuation for analyzing deterministic eavesdropping, while fast fluctuations account for probabilistic eavesdropping.

**Table 5** Classification of turbulence strength [94]

| Turbulence strength | $C_n^2$ (m$^{-2/3}$) |
| --- | --- |
| Weak | $< 10^{-17}$ |
| Moderate | $(10^{-17}, 10^{-13})$ |
| Strong | $> 10^{-13}$ |

Our model assumes a point-to-point outdoor THz wireless channel, as shown in Fig. 20(a). Alice transmits information to Bob via a LoS channel through absorbing and scattering turbulence, described by Eq. (18). An eavesdropper (Eve), positioned near the beam, aims to capture data through a NLoS path. For a signal transmitted along the *x*-axis from Alice at (0,0) to Bob at (d,0) with Eve at (*x*, *y*), the NLoS channel gain $G_{NLoS}$ [219] can be obtained by Eq. (19).

We evaluated the model's accuracy using a 625 GHz wireless channel through emulated atmospheric turbulence in a weather chamber [62]. Turbulence was generated by introducing airflows at different temperatures and speeds. Turbulence strength varied from $3.5 \times 10^{-11}$ m$^{-2/3}$ to $2.3 \times 10^{-9}$ m$^{-2/3}$, corresponding to maximum Rytov variances. The agreement between theoretical and experimental data for a 1m channel, confirms the applicability of Eq. (18). Fig. 24 shows attenuation across the THz spectrum for different turbulence strengths, aligning with spectral absorption trends but offset by turbulence effects. Channels at 140, 220, 340, and 675 GHz were modeled for eavesdropping vulnerability over a 1 km distance.

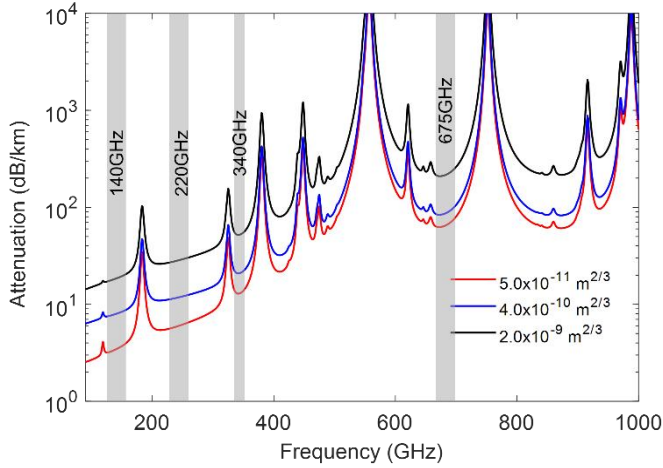

**Figure 24 Attenuation due to atmospheric turbulence with different strengths. (pressure *P* = 1013 hPa, humidity RH = 20%, channel distance *d* =1km).**

Beam divergence $α_A$ = 20 mrad accounts for small misalignments at THz frequencies. Fig. 25(a) shows channel gain vs. turbulence strength up to $C_n^2 = 1.0 \times 10^{-10}$ m$^{-2/3}$. Solid lines represent $G_{LoS}$ for Bob, while dashed lines show $G_{NLoS}$ for Eve. Close proximity of Eve to the LoS path increases eavesdropping risk, especially when turbulence strength exceeds $C_n^2 = 1.1 \times 10^{-11}$ m$^{-2/3}$. We use Wyner's secrecy capacity metric (Eq. (20)) to evaluate deterministic eavesdropping risk. Fig. 25(b) shows the secrecy capacity for Eve's x-position varying from 0 m to 1 km with y = 50 m. When Eve is between 70 m and 460 m, the secrecy capacity is 0 Gbps, indicating an 'insecure region'. Fig. 25(c) illustrates the secrecy capacity as y-position changes at x = 500 m. For y ≤ 50 m, the region is insecure. Beyond 50 m, the secrecy capacity increases, reaching a maximum (MSC). This suggests that increasing the minimum distance from Eve to the LoS path can mitigate eavesdropping risks from atmospheric turbulence. Fig. 25(d) shows the 2-D secrecy capacity distribution for Eve's arbitrary positions, with the color bar denoting the secrecy capacity in Gbps. The brightest color represents an MSC of 67 Gbps, while dark blue indicates the insecure region (0 Gbps). Horizontal and vertical white lines show the secrecy capacity evolution versus x- and y-positions of Eve, as in Fig. 25(b) and (c).

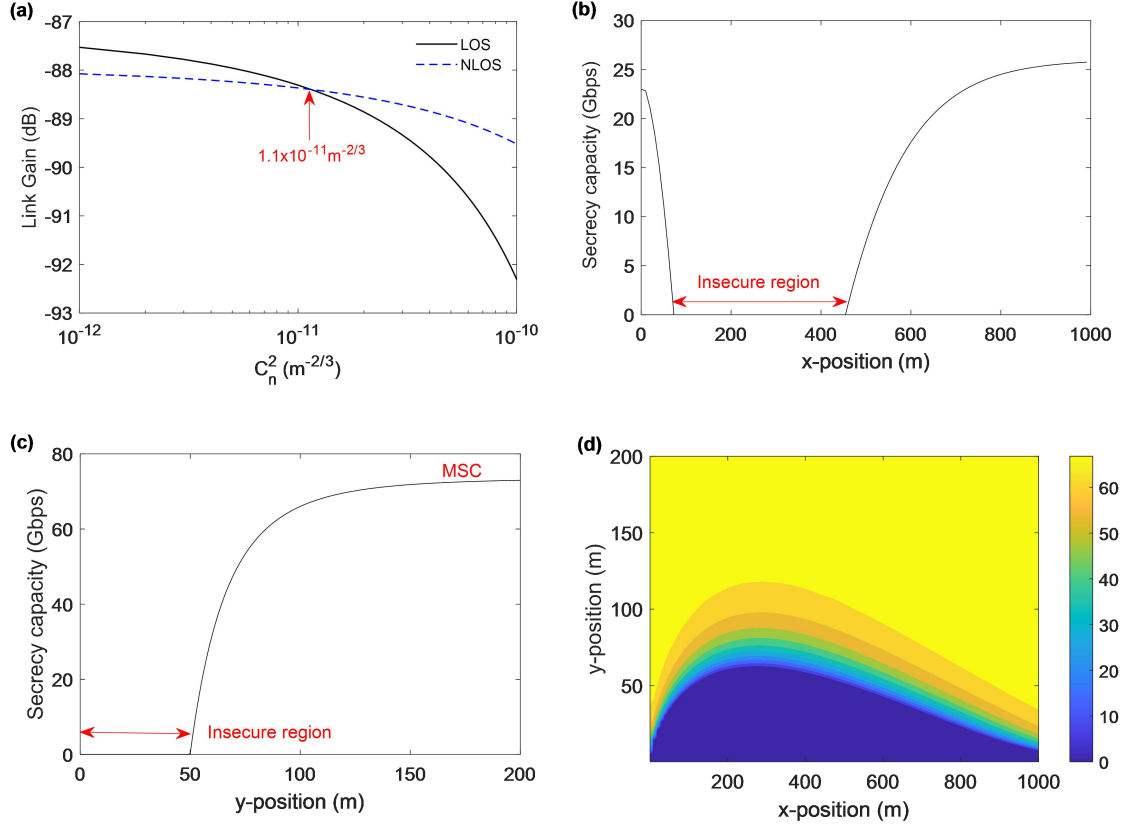

Figure 25 (a) Evolution of channel gain received by Bob (LoS) and Eve (NLoS) versus turbulence strength when Eve is located at (500m, 50m); (b) Evolution of channel secrecy capacity distribution versus *x*-position of Eve (*y* = 50m); (c) Evolution of channel secrecy capacity distribution versus *y*-position of Eve (*x* = 500m); (d) Secrecy capacity distribution for 2-D positions of Eve with a unit of Gbps in the color bar.

To understand the dependence of eavesdropping risk on carrier frequencies, we calculate the secrecy capacity at 140, 220, 340, and 675 GHz as shown in Fig. 26(a). As carrier frequencies increase, MSC decreases significantly, and the insecure region expands due to more scattering and higher gaseous attenuation. At 675 GHz, the secrecy capacity reaches 0 Gbps, making the whole region insecure. Higher frequencies thus face more serious multipath scattering and eavesdropping risks. Fig. 26(b) shows that for a 340 GHz channel, increasing turbulence strength from $C_n^2 = 10^{-12}$ m$^{-2/3}$ to $10^{-10}$ m$^{-2/3}$ decreases MSC and expands the insecure region. Fig. 26(c) indicates that increasing divergence angle $\alpha_A$ from 25 mrad to 35 mrad expands the insecure region and reduces MSC from 45 Gbps to 22 Gbps. Assuming Eve has complete channel state information (CSI) and computational capabilities, Fig. 26(d) estimates secrecy capacity for different SNR values and FOV angles. The insecure region enlarges with Eve's SNR decreasing from 6 dB to 0 dB, though MSC remains unaffected. Changes in FOV angle (5° or 20°) do not significantly affect capacity, suggesting this is not an effective strategy for reducing eavesdropping risks.

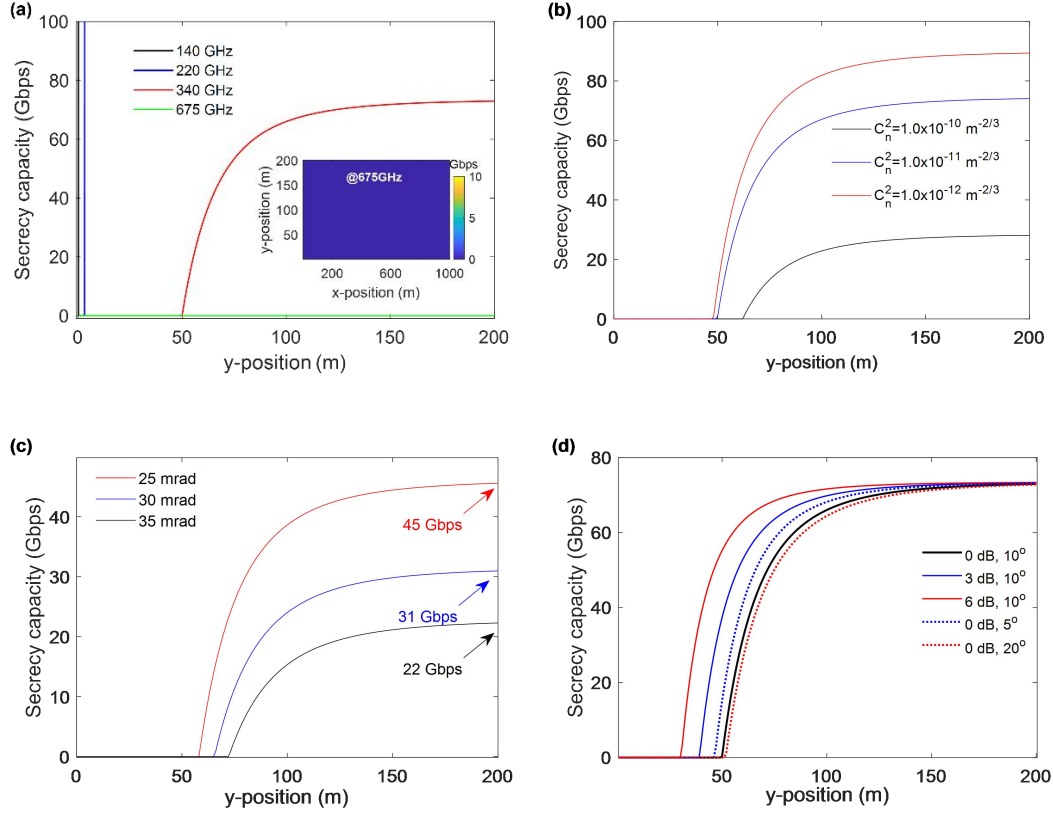

**Figure 26** Variation of secrecy capacity with respect to the *y*-position of Eve under different (a) carrier frequencies, (b) turbulence strengths, (c) divergence angles, and (d) receiver sensitivities and FOV angle for Eve. Inset of (a): secrecy capacity distribution for 2-D positions of Eve with a unit of Gbps in the color bar.

Due to scattering-induced scintillation effects, received power and channel gain fluctuate [225], requiring the use of outage probability to assess eavesdropping risk. The outage probability is the likelihood that the instantaneous secrecy capacity falls below a target rate $R$, calculated as $P_0(R) = P_r\{C_s < R\}$ [226] where G is the solution for $C_s = R$ and $G_{\text{LOS}}$ is the instantaneous LoS channel gain. The probability density function of $G_{\text{LOS}}$ [94] is given by

$$f_{\text{LOS}}(G_{\text{LOS}}) = \frac{1}{G_{\text{LOS}}\sqrt{2\pi\sigma_r^2}} \cdot \exp\left[-\frac{\left(\log\left(G_{\text{LOS}}/\overline{G_{\text{LOS}}}\right) - \left\langle\log\left(G_{\text{LOS}}/\overline{G_{\text{LOS}}}\right)\right\rangle\right)^2}{2\sigma_r^2}\right] \quad (23)$$

Using Eq. (19), we analyze outage probability for $R$=10 Gbps, common in THz wireless channels [227]. As shown in Fig. 27(a), higher frequencies like 675 GHz exhibit high outage probability, indicating greater eavesdropping risk, while 140 GHz and 220 GHz channels remain secure. For 340 GHz, the risk becomes significant beyond 33m. Fig. 27(b) shows that stronger turbulence increases outage probability for 340 GHz channels. In Fig. 27(c), increasing beam divergence also raises the risk. Fig. 27(d) illustrates that reducing Eve's SNR increases the insecure region, while varying FOV angle has minimal impact, suggesting that manipulating SNR is a more effective strategy for mitigating eavesdropping.

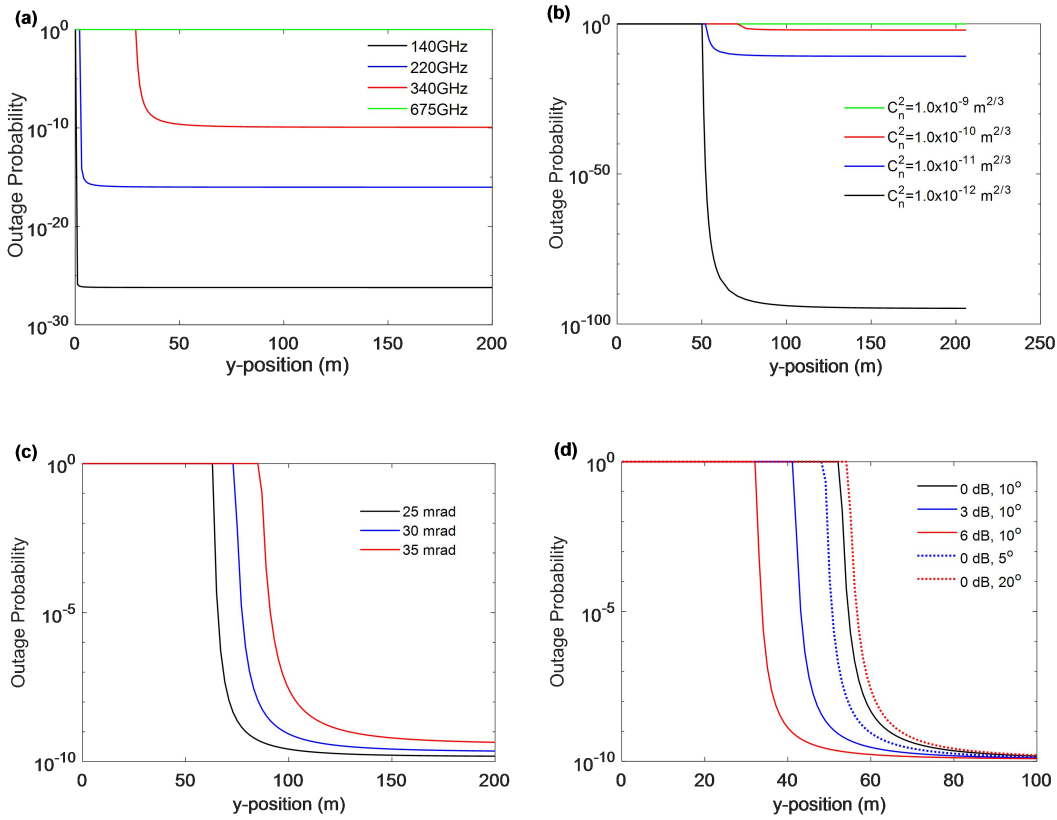

**Figure 27 Variation of outage probability with respect to the y-position of Eve for different (a) carrier frequencies, (b) turbulence strengths, (c) divergence angles, and (d) receiver sensitivities and FOV angle of Eve.**

### 4.4 Comparative analysis of different weather conditions

For all weather conditions examined (rain, snow, and atmospheric turbulence), higher carrier frequencies generally lead to increased vulnerability to eavesdropping. This is due to greater signal attenuation and scattering at higher frequencies. For example, the 340 GHz and 675 GHz channels consistently showed larger insecure regions and lower secrecy capacities compared to 140 GHz and 220 GHz channels. This trend was especially pronounced for atmospheric turbulence, where the 675 GHz channel became completely insecure across the entire region.

The scattering of THz signals plays a major role in enabling eavesdropping for all weather conditions studied. Interestingly, wet snow generally showed better secrecy performance than rain at equivalent precipitation rates. This is likely due to differences in particle size distribution and scattering properties. Dry snow, however, posed greater eavesdropping risks than both rain and wet snow due to its higher particle density and stronger scattering effects.

Atmospheric turbulence also induces scattering, with stronger turbulence increasing eavesdropping risks.

Future researches should prioritize the development of adaptive communication techniques that can dynamically adjust link parameters in response to changing weather conditions. For instance, lower frequency bands should be favored in environments with frequent heavy rain or snow, as they exhibit better resilience against attenuation and scattering. Furthermore, in regions with significant atmospheric turbulence, innovative beamforming and alignment strategies must be developed to mitigate beam divergence and maintain secure links. Future research should also explore the integration of real-time weather monitoring systems with THz communication networks. This integration can enable proactive adjustments to transmission parameters based on current and forecasted weather conditions, enhancing the overall robustness and security of the communication system. Additionally, the exploration of hybrid communication systems that combine THz links with other frequency bands could provide a more resilient solution to weather-induced eavesdropping risks.

## 5. Future Research Directions

As THz communication technology advances and moves towards practical outdoor applications, several key areas require further research to address the challenges posed by atmospheric conditions and security concerns:

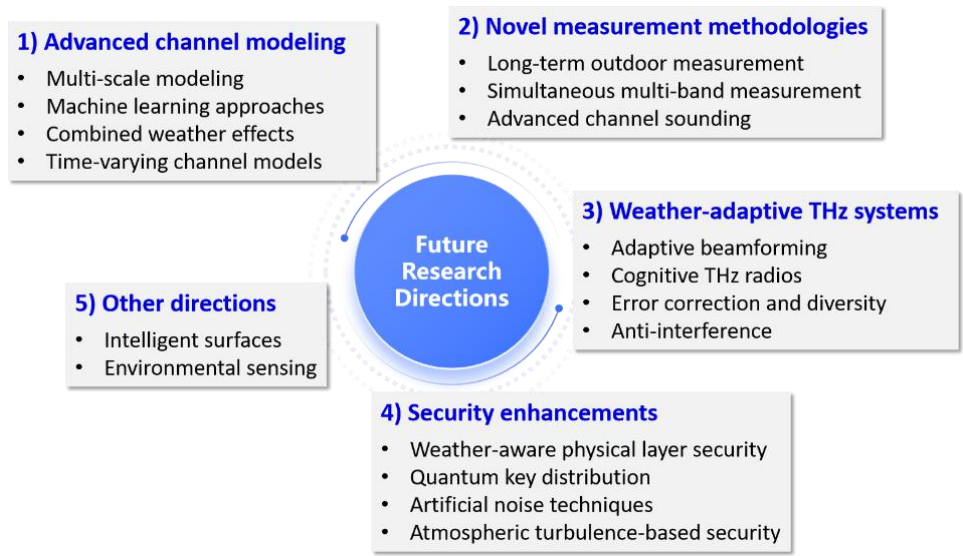

Figure 28 Outline of future research directions

1) Advanced channel modeling

Future research in THz communications should focus on developing comprehensive, multi-scale channel models that accurately capture the complex interplay of atmospheric effects on signal propagation. These models need to simultaneously account for large-scale phenomena, such as water vapor absorption, and small-scale scattering from particles like rain droplets or snowflakes. An integrated approach combining radiative transfer models for macroscopic effects with detailed electromagnetic simulations of microscopic interactions would provide a more holistic understanding of THz channel behavior across different spatial scales [228].

To address the inherent complexity and dynamism of atmospheric conditions, AI and machine learning techniques offer promising solutions. Deep neural networks trained on extensive datasets of atmospheric parameters and THz channel characteristics could learn to model complex, non-linear relationships more effectively than traditional analytical approaches [229, 230]. These AI-driven models could be particularly valuable in scenarios involving multiple interacting weather effects, which are common in real-world environments but often overlooked in current studies that focus on isolated phenomena. Furthermore, the temporal variability of atmospheric conditions necessitates the development of dynamic channel models. Incorporating time-series analysis techniques or state-space modeling could enable these models to track the evolution of channel parameters over short time scales. Such capabilities would be crucial for designing adaptive communication protocols that can respond in real-time to changing weather conditions, ensuring robust performance of THz channels in variable environments. The application of machine learning in THz communications extends beyond channel modeling. Researchers have successfully applied machine learning techniques to various aspects of the physical layer, including modulation recognition [231, 232], channel identification [233, 234], encoding and decoding [235, 236], channel estimation [237], and equalization [238, 239]. These advancements provide valuable insights for future research on physical layer security of THz channels.

The ultimate goal of these advanced modeling efforts should be to create comprehensive frameworks that can account for the combined impact of various weather phenomena and their interactions on THz channel propagation. This may involve developing sophisticated statistical models that capture correlations between different atmospheric parameters, or physics-based simulations that model the interplay of various weather effects [50, 240]. By providing a more realistic representation of THz channel behavior in diverse and dynamic atmospheric conditions, these models will play a vital role in the design and optimization of future THz communication systems, enabling them to maintain reliable performance across a wide range of weather scenarios.

2) **Novel measurement methodologies**

To advance our understanding of THz channel propagation in atmospheric conditions, new and innovative measurement techniques are essential. We suggest that future research should focus on three key areas:

① Long-term outdoor measurements in diverse geographical locations are crucial for capturing the full range of weather conditions and their impacts on THz channels. While controlled laboratory experiments provide valuable insights, they cannot fully replicate the complexity of real-world environments. Continuous monitoring over extended periods (e.g., months or years) will provide a wealth of data to validate and refine channel models, as well as reveal potential unforeseen effects that may only become apparent over time. For example, researchers at the Norwegian University of Science and Technology (NTNU) have initiated a long-term outdoor measurement campaign using several mm-Wave links to study seasonal variations in channel performance [241]. Such studies serve as a model for future THz measurement campaigns.

② Concurrent measurements across multiple frequency bands are essential for understanding frequency-dependent weather effects. By simultaneously characterizing channel behavior at mm-Wave, THz, and even infrared frequencies under the same atmospheric conditions, researchers can gain valuable insights into how different weather phenomena impact various parts of the spectrum [38, 43]. A recent study by researchers at New York University compared the performance of 28 GHz, 140 GHz, and 1.5 THz channels under various weather conditions, revealing important differences in their susceptibility to atmospheric effects [242]. Expanding on this approach with more comprehensive multi-band measurement campaigns will be crucial for optimizing frequency selection and diversity techniques in future communication systems.

③ Developing wideband, high-resolution channel sounders specifically designed for THz frequencies in outdoor environments is a critical area for future work. These systems must overcome challenges such as high path loss, atmospheric absorption, and the need for precise timing synchronization at THz frequencies. Future research should explore novel architectures, such as photonics-based systems or distributed channel sounding networks, to achieve the necessary bandwidth, dynamic range, and measurement speed for characterizing rapidly varying THz channels. Additionally, integrating environmental sensors (e.g., temperature, humidity, particle counters) with THz channel sounders could provide correlated data on atmospheric conditions and channel characteristics, enabling more comprehensive analysis of weather effects.

By combining these three approaches, we can develop a more comprehensive and nuanced understanding of THz channel propagation in diverse atmospheric conditions. This integrated approach to measurement will be instrumental in developing robust, weather-resilient THz communication systems for future applications.

### 3) Weather-adaptive THz systems

As THz communication systems move towards real-world deployment, developing weather-adaptive technologies becomes crucial for maintaining reliable performance under varying atmospheric conditions. Future research in this area should focus on four interconnected aspects: adaptive beamforming, cognitive radio systems, advanced coding and diversity schemes, and anti-interference technologies.

Adaptive beamforming techniques represent a promising approach to mitigate weather-induced propagation effects. We should investigate methods to dynamically adjust beam patterns in response to atmospheric conditions. Algorithms could be developed to compensate for beam wandering and broadening caused by atmospheric turbulence [55]. These techniques could leverage real-time channel state information and atmospheric sensing to optimize beam shape and direction, potentially utilizing large-scale antenna arrays or reconfigurable intelligent surfaces. The challenge lies in developing fast and accurate algorithms that can respond to rapidly changing atmospheric conditions while maintaining system stability.

Complementing adaptive beamforming, cognitive THz radio systems that can sense atmospheric conditions and adapt transmission parameters in real-time are another critical area for future research. These systems could integrate environmental sensors and channel quality metrics to make intelligent decisions about frequency selection, transmission power, modulation scheme, and other parameters [229]. During heavy rainfall, the system might switch to a lower frequency or a more robust modulation scheme. Key research challenges include developing fast and accurate sensing mechanisms, creating efficient decision-making algorithms, and designing flexible hardware architectures that can rapidly reconfigure transmission parameters.

To further enhance the reliability of THz links in adverse weather conditions, advanced coding and diversity schemes tailored to the unique characteristics of weather-impaired THz channels are essential. The research community should focus on designing error correction codes capable of handling the bursty error patterns often encountered in THz channels affected by atmospheric phenomena [243]. Additionally, novel diversity techniques that exploit the wide bandwidth available in THz systems should be explored. Frequency diversity schemes could strategically allocate data across multiple sub-bands to mitigate frequency-selective fading caused by water vapor absorption lines. Spatial diversity techniques using massive MIMO configurations could also be investigated to combat signal fading and beam misalignment in turbulent atmospheric conditions.

Anti-interference technologies offer promising solutions for mitigating the impact of adverse weather on THz communications. Spread spectrum techniques, such as Direct Sequence Spread Spectrum (DSSS), have shown potential in improving the bit error rate performance of wireless channels under turbulent atmospheric conditions [244]. By spreading the signal over a wider bandwidth, these techniques can enhance resistance to narrowband interference and multipath fading induced by atmospheric scattering. Frequency modulation, particularly Orthogonal Frequency Division Multiplexing (OFDM), presents another effective approach to combat frequency-selective fading in THz channels affected by weather. Adaptive OFDM schemes have demonstrated significant improvements in capacity and reliability of THz channels under varying atmospheric conditions [245]. By dividing the channel into multiple narrowband subcarriers, OFDM can adapt to frequency-dependent atmospheric attenuation, maintaining link quality even in challenging weather scenarios. Domain adaptation techniques, leveraging machine learning approaches, show great promise in enabling THz communication systems to adapt to changing atmospheric conditions in real-time. Reinforcement learning-based approaches can be proposed to dynamically adjust beamforming and power allocation in THz networks under varying atmospheric conditions. These methods can optimize transmission parameters based on current channel conditions, potentially mitigating the effects of rain, snow, or turbulence more effectively than traditional static approaches.

4) Security enhancements

With THz communication systems moving towards widespread outdoor deployment, ensuring robust security becomes increasingly critical. The unique propagation characteristics of THz waves in the atmosphere present both challenges and opportunities for enhancing link security. Future research should focus on the following areas:

Adaptive security protocols that can respond to changing atmospheric conditions and potential eavesdropping risks are crucial for THz communications. We should explore techniques to dynamically adjust the broadcast range based on atmospheric absorption, as demonstrated by Fang et al. [245]. This approach exploits strong water vapor absorption lines in the THz band to create a "secure zone" where communication is possible, while signals beyond this range are too attenuated for potential eavesdroppers to detect. Future work should integrate real-time weather data and propagation models to optimize these security protocols under varying atmospheric conditions.

Quantum cryptography using THz frequencies in outdoor environments is an exciting area for future research. While most current quantum key distribution (QKD) systems operate in the optical domain, the THz band offers potential benefits such as reduced background noise and compatibility with electronic systems [246]. We should explore adapting existing QKD protocols for THz frequencies and developing new protocols that leverage unique THz propagation characteristics. Key challenges include mitigating atmospheric turbulence and absorption effects on quantum state preservation, and developing efficient THz single-photon sources and detectors suitable for outdoor use.

Artificial noise techniques represent another promising approach to enhance THz channel security. Future work should investigate methods to introduce controlled interference that degrades eavesdropper performance without significantly impacting legitimate users. Researchers could explore shaping the spatial and spectral properties of artificial noise to exploit the directional nature of THz channels and frequency-selective atmospheric absorption [247]. Multiple antenna elements could be used to create null spaces in the artificial noise pattern at the intended receiver's location while maintaining high interference levels in other directions. Additionally, the wide THz bandwidth could be leveraged to develop frequency-hopping artificial noise schemes that are difficult for eavesdroppers to track.

Lastly, exploiting the unique properties of atmospheric turbulence for security enhancement is an intriguing direction for future research. While turbulence is generally considered a challenge for THz communications, it could potentially be leveraged as a natural source of randomness for security purposes. We could explore using the rapidly varying channel state induced by turbulence as a shared secret key between the transmitter and legitimate receiver. Another possibility is developing protocols that exploit the spatial decorrelation of turbulence-induced fading to create a security advantage for the intended receiver over a potential eavesdropper at a different location. These approaches would require developing accurate models of THz channel propagation through turbulent atmospheres and designing robust signal processing techniques to extract security-relevant information from turbulence-induced channel variations.

### 5) Other directions

Beyond the specific areas mentioned earlier, several additional research directions hold significant promise for advancing THz communications in outdoor environments.

A critical approach is the development of high-power terahertz radiation sources and highly sensitive receivers. These components are fundamental to overcoming the atmospheric attenuation challenges discussed in this manuscript. High-power sources can maintain channel strength over longer distances in adverse weather conditions, while sensitive receivers can improve channel detection. Research should focus on novel materials and device architectures for efficient THz wave generation and detection. Advances in semiconductor technology, such as improved quantum cascade lasers or resonant tunneling diodes for generation, and enhanced mixer or direct detection technologies for reception, could significantly boost THz communication capabilities. These high-performance components are crucial for realizing practical, long-range THz communications that can operate reliably in various atmospheric conditions.

Engineered surfaces offer exciting possibilities for manipulating THz channel propagation and mitigating weather effects. Intelligent surfaces, consisting of sub-wavelength structures, can be designed to control the reflection, refraction, and absorption of THz waves in ways not possible with conventional materials [248, 249]. Future research should focus on developing adaptive surfaces that can dynamically adjust their properties in response to changing weather conditions. These surfaces could potentially reduce the impact of rain or snow on THz channel propagation, or even exploit weather effects to enhance channel performance. However, several challenges need to be addressed, including designing surfaces that can operate over wide bandwidths, developing fabrication techniques suitable for large-scale outdoor deployment, and creating control systems that can rapidly reconfigure surface properties in real-time. By overcoming these challenges, engineered surfaces could become a powerful tool for enhancing the reliability and performance of outdoor THz communication systems.

Leveraging the sensitivity of THz waves to atmospheric conditions presents an intriguing opportunity for developing dual-use systems that combine communication and environmental monitoring capabilities. The same characteristics that make THz propagation challenging in adverse weather also make it highly sensitive to atmospheric composition and state. Researchers could explore ways to extract information about temperature, humidity, and particulate concentrations from received THz signals in communication links [230]. This approach could lead to the development of systems that provide both high-speed wireless connectivity and distributed atmospheric sensing capabilities. Future work in this area might involve developing signal processing algorithms to separate environmental effects from communication data, designing multi-functional THz transceivers that can switch between communication and sensing modes, and creating data fusion techniques to combine information from multiple THz links for improved environmental monitoring accuracy. Such dual-use systems could not only enhance the efficiency of THz spectrum utilization but also provide valuable environmental data for weather forecasting, air quality monitoring, and climate studies.

These research directions represent innovative approaches to addressing the challenges and exploiting the unique properties of THz waves in outdoor environments. By pursuing these avenues alongside the previously mentioned areas, researchers can contribute to a more comprehensive and multifaceted advancement of THz communication technology. The integration of these diverse research directions will be crucial in realizing the full potential of THz communications for next-generation wireless networks and beyond.

## 6. Conclusions

As research in THz communications progresses, it is becoming increasingly clear that a thorough understanding of THz channel propagation characteristics and security properties is paramount for the successful deployment of future THz communication systems. This understanding will be crucial in addressing the challenges and fully exploiting the potential of the THz spectrum in the next generation of wireless networks. This review article has examined the propagation characteristics and security implications of terahertz (THz) channels in atmospheric conditions, with a focus on the impact of various weather phenomena.

THz channels are significantly affected by atmospheric conditions, primarily through absorption, scattering, and refractive effects. Water vapor absorption creates distinct transmission windows in the THz spectrum, while particles such as rain droplets and snowflakes can cause scattering. Atmospheric turbulence induces scintillation effects, which become more pronounced at higher frequencies. A variety of techniques, including THz time-domain spectroscopy, vector network analyzer methods, and channel sounders, have been developed to characterize THz channels in atmospheric conditions. Modeling approaches range from deterministic methods based on electromagnetic theory to statistical models that capture the random nature of atmospheric effects. Hybrid models offer a promising approach to balance accuracy and computational efficiency. The impact of weather conditions on THz channels is highly frequency-dependent and can vary significantly across different atmospheric phenomena. Rain and snow, for instance, pose substantial challenges due to their strong scattering effects, particularly at higher frequencies. Atmospheric turbulence, while less severe in terms of overall attenuation, introduces complex fading characteristics that can significantly impact system performance. These findings underscore the need for adaptive and robust communication strategies that can dynamically respond to changing atmospheric conditions to maintain reliable THz links.

Furthermore, investigations into the security aspects of THz communications in adverse weather conditions have highlighted both challenges and opportunities. While weather-induced scattering can potentially increase vulnerability to eavesdropping, it also opens up possibilities for novel security protocols that leverage the unique propagation characteristics of THz waves in the atmosphere. The development of weather-aware security measures will be critical in ensuring the confidentiality and integrity of future THz communication systems. As research in this field advances, a multidisciplinary approach combining expertise in electromagnetics, atmospheric science, and information security will be essential to address these complex challenges and realize the full potential of THz communications in outdoor environments.

It is important to acknowledge that this review has primarily focused on the effects of natural atmospheric phenomena on THz communications. However, human-induced factors also play a significant role in real-world deployments. Electromagnetic interference from other wireless devices and systems operating in nearby frequency bands can potentially disrupt THz communications, necessitating research into interference mitigation techniques. Urban infrastructure, such as buildings, can cause complex reflection, diffraction, and scattering of THz signals, requiring sophisticated channel models that account for these effects. Additionally, industrial emissions and pollutants may alter atmospheric composition, potentially affecting THz signal propagation in ways that are not yet fully understood. Future research should aim to incorporate these human-induced factors into comprehensive THz channel models and system designs.

## Acknowledgement

We appreciate the support by the National Science Foundation of China (62071046, 62471033), the U.S. National